\documentclass[11pt]{article}

\usepackage{latexsym,amsmath}

\newcounter{subequation}[equation]
\makeatletter

\def\bcite{\@ifnextchar [{\@tempswatrue\@bcitex}{\@tempswafalse\@bcitex[]}}
\def\@bcitex[#1]#2{\if@filesw\immediate\write\@auxout{\string\citation{#2}}\fi
  \let\@bcitea\@empty
  \@bcite{\@for\@bciteb:=#2\do
    {\@bcitea\def\@bcitea{,\penalty\@m\ }%
     \def\@tempa##1##2\@nil{\edef\@bciteb{\if##1\space##2\else##1##2\fi}}%
     \expandafter\@tempa\@bciteb\@nil
     \@ifundefined{b@\@bciteb}{{\reset@font\bf ?}\@warning
       {Citation `\@bciteb' on page \thepage \space undefined}}%
     \hbox{\csname b@\@bciteb\endcsname}}}{#1}}
\def\@bcite#1#2{{#1\if@tempswa , #2\fi}}

\def\thesubequation{\theequation\@alph\c@subequation}
\def\@subeqnnum{{\rm (\thesubequation)}}
\def\slabel#1{\@bsphack\if@filesw {\let\thepage\relax
   \xdef\@gtempa{\write\@auxout{\string
      \newlabel{#1}{{\thesubequation}{\thepage}}}}}\@gtempa
   \if@nobreak \ifvmode\nobreak\fi\fi\fi\@esphack}
\def\subeqnarray{\stepcounter{equation}
\let\@currentlabel=\theequation\global\c@subequation\@ne
\global\@eqnswtrue
\global\@eqcnt\z@\tabskip\@centering\let\\=\@subeqncr
$$\halign to \displaywidth\bgroup\@eqnsel\hskip\@centering
  $\displaystyle\tabskip\z@{##}$&\global\@eqcnt\@ne
  \hskip 2\arraycolsep \hfil${##}$\hfil
  &\global\@eqcnt\tw@ \hskip 2\arraycolsep
  $\displaystyle\tabskip\z@{##}$\hfil
   \tabskip\@centering&\llap{##}\tabskip\z@\cr}
\def\endsubeqnarray{\@@subeqncr\egroup
                     $$\global\@ignoretrue}
\def\@subeqncr{{\ifnum0=`}\fi\@ifstar{\global\@eqpen\@M
    \@ysubeqncr}{\global\@eqpen\interdisplaylinepenalty \@ysubeqncr}}
\def\@ysubeqncr{\@ifnextchar [{\@xsubeqncr}{\@xsubeqncr[\z@]}}
\def\@xsubeqncr[#1]{\ifnum0=`{\fi}\@@subeqncr
   \noalign{\penalty\@eqpen\vskip\jot\vskip #1\relax}}
\def\@@subeqncr{\let\@tempa\relax
    \ifcase\@eqcnt \def\@tempa{& & &}\or \def\@tempa{& &}
      \else \def\@tempa{&}\fi
     \@tempa \if@eqnsw\@subeqnnum\refstepcounter{subequation}\fi
     \global\@eqnswtrue\global\@eqcnt\z@\cr}
\let\@ssubeqncr=\@subeqncr
\@namedef{subeqnarray*}{\def\@subeqncr{\nonumber\@ssubeqncr}\subeqnarray}
\@namedef{endsubeqnarray*}{\global\advance\c@equation\m@ne%
                           \nonumber\endsubeqnarray}

\makeatother

\DeclareFontFamily{OT1}{rsfs10}{}
\DeclareFontShape{OT1}{rsfs10}{m}{n}{ <-> rsfs10 }{}
\DeclareMathAlphabet{\mathscript}{OT1}{rsfs10}{m}{n}

\numberwithin{equation}{section}

\newcommand{\ns}{\normalsize}
\newcommand{\pt}{\partial}

\newcommand{\be}{\begin{equation}}
\newcommand{\ee}{\end{equation}}
\newcommand{\nn}{\nonumber}
\newcommand{\bea}{\begin{eqnarray}}
\newcommand{\eea}{\end{eqnarray}}
\newcommand{\bsea}{\begin{subeqnarray}} 
\newcommand{\esea}{\end{subeqnarray}}

\newcommand{\orbav}[1]{\big<{#1}\big>_{11}}
\newcommand{\tr}{\textrm{tr}}

\def\a{\alpha}
\def\b{\beta}
\def\g{\gamma}
\def\c{\chi}
\def\d{\delta}
\def\e{\epsilon}

\def\z{\psi}

\def\k{\kappa}
\def\l{\lambda}
\def\m{\mu}
\def\n{\nu}
\def\o{\omega}
\def\p{\pi}

\def\r{\rho}
\def\s{\sigma}

\def\t{\tau}

\def\x{\xi}
\def\z{\zeta}
\def\w{\wedge}
\def\D{\Delta}

\def\G{\Gamma}
\def\J{\Psi}

\def\O{\Omega}

\def\cA{{\cal A}}
\def\cF{{\cal F}}
\def\cM{{\cal M}}
\def\cK{{\cal K}}
\def\cC{{\cal C}}
\def\cP{{\cal P}}
\def\cL{{\cal L}}

\def\vb{\bar{v}}
\def\ub{\bar{u}}
\def\xib{\bar{\xi}}

\begin{document}


\begin{titlepage}

\vspace{-3cm}

\title{
   \hfill{\ns UPR-804T, PUPT-1793, CERN-TH/98-178, Imperial/TP/97-98/52\\}
   \hfill{\ns hep-th/9806051\\[.5cm]}
   {\LARGE Heterotic M--theory in Five Dimensions}}
\author{
   Andr\'e Lukas$^1$
      \setcounter{footnote}{0}\thanks{Supported in part by Deutsche
          Forschungsgemeinschaft (DFG).}~~,
   Burt A.~Ovrut$^1$
      \setcounter{footnote}{3}\thanks{Supported in part by a Senior 
          Alexander von Humboldt Award.}~~,
   K.S. Stelle$^2$ and Daniel Waldram$^3$\\[0.5cm]
   {\ns $^1$Department of Physics, University of Pennsylvania} \\
   {\ns Philadelphia, PA 19104--6396, USA}\\[0.3cm]
   {\ns $^2$The Blackett Laboratory, Imperial College, London SW7 2BZ, UK}\\ 
   {\ns and}\\
   {\ns TH Division, CERN, CH-1211 Geneva 23, Switzerland}\\[0.3cm]
   {\ns $^3$Department of Physics, Joseph Henry Laboratories,}\\ 
   {\ns Princeton University, Princeton, NJ 08544, USA}}
\date{}

\maketitle

\begin{abstract}
We derive the five--dimensional effective action of strongly coupled
heterotic string theory for the complete $(1,1)$ sector of the theory
by performing a reduction, on a Calabi--Yau three--fold, of M--theory
on $S^1/Z_2$. A crucial ingredient for a consistent truncation is a
non--zero mode of the antisymmetric tensor field strength which arises
due to magnetic sources on the orbifold planes. The correct effective theory
is a {\it gauged} version of five--dimensional $N=1$ supergravity
coupled to Abelian vector multiplets, the universal hypermultiplet and
four--dimensional boundary theories with gauge and gauge matter
fields. The gauging is such that the dual of the four--form field strength
in the universal multiplet is charged under a particular linear combination
of the Abelian vector fields. In addition, the theory has potential
terms for the moduli in the bulk as well as on the boundary. Because
of these potential terms, the supersymmetric ground state of the
theory is a multi--charged BPS three--brane domain wall, which we
construct in general. We show that the five--dimensional theory
together with this solution provides the correct starting point for
particle phenomenology as well as early universe cosmology. As an
application, we compute the four--dimensional $N=1$ supergravity
theory for the complete $(1,1)$ sector to leading nontrivial order by
a reduction on the domain wall background. We find a correction to the
matter field K\"ahler potential and threshold corrections to the gauge
kinetic functions.  
\end{abstract}

\thispagestyle{empty}

\end{titlepage}


\section{Introduction}


One of the phenomenologically most promising corners of the M--theory
moduli space, in addition to the weakly coupled heterotic string, is
the point described at low-energy by eleven-dimensional supergravity
on the orbifold $S^1/Z_2$ due to Ho\v rava and
Witten~\cite{hw1,hw2}. This theory gives the strongly coupled limit of
the heterotic string with, in addition to the bulk supergravity, two
sets of $E_8$ gauge fields residing one on each of the two
ten--dimensional fixed hyperplanes of the orbifold. It has been
shown~\cite{w} that this theory has phenomenologically interesting
compactifications on deformed Calabi--Yau three--folds times the
orbifold to four dimensions. Matching the 11--dimensional Newton
constant $\k$, the Calabi--Yau volume and the orbifold radius to the
known values of the Newton constant and the grand unification coupling
and scale leads to an orbifold radius which is about an order of
magnitude or so larger than the two other scales~\cite{w,bd}. This
suggests that, near this ``physical'' point in moduli space, the
theory appears effectively five--dimensional in some intermediate
energy regime. 

In a previous paper~\cite{losw} we have derived this five--dimensional
effective theory for the universal fields for the first time by
directly reducing Ho\v rava--Witten theory on a Calabi--Yau three--fold.
We showed that a non--zero mode of the antisymmetric tensor field
strength has to be included for a consistent reduction from eleven to
five dimensions and that the correct five--dimensional effective
theory of strongly coupled heterotic string is given by a gauged
version of five--dimensional supergravity. A reduction of pure
eleven--dimensional supergravity on a Calabi--Yau
three--fold~\cite{CYred}, on the other hand, leads to a non--gauged
version of five--dimensional supergravity. Therefore, while this
provides a consistent low--energy description of M--theory on a smooth
manifold it is not the correct effective theory for M--theory on
$S^1/Z_2$. The necessary additions are chiral four--dimensional
boundary theories with potential terms for the bulk moduli and, most
importantly, the aforementioned non--zero mode, living solely in the
Calabi-Yau, which leads to the gauging of the bulk supergravity. As
pointed out in ref.~\cite{losw} this theory is the correct starting
point for strongly coupled heterotic particle phenomenology as well as
early universe cosmology if the theory indeed undergoes a
five--dimensional phase as suggested by the data. Moreover, we have
shown that contact with four--dimensional physics should not be made
using flat space--time but rather a domain--wall solution as the
background configuration. This domain wall arises as a BPS state of
the five--dimensional theory~\cite{losw} and its existence is
intimately tied to the gauging of the theory. A reduction to four
dimensions on this domain wall has been performed in~\cite{paschos} to
lowest non--trivial order. The result agrees with ref.~\cite{low1}
where the complete four--dimensional effective action to that order
has been derived directly from eleven dimensions. This is not
surprising given that all reductions have been carried out
consistently. 

Various other aspects of the Ho\v rava--Witten description of strongly coupled
heterotic string theory have been addressed in the literature such as
the structure of the four--dimensional effective action, its relation
to 10--dimensional weakly coupled heterotic string, gaugino condensation,
and anomaly cancelation~[\bcite{low1}\,--\,\bcite{bkl}]. Aspects of
five--dimensional physics motivated by Ho\v rava--Witten theory and related
to particle phenomenology have been discussed in
ref.~\cite{bd,noy,sharpe,pes}. In refs.~\cite{lo,benakli,low4}
five--dimensional early universe M--theory cosmology have been
investigated. Recently, aspects of five--dimensional physics have also
been discussed in ref.~\cite{elpp}. 

The main purpose of the present paper is to generalize the result of
ref.~\cite{losw} to include the full $(1,1)$ sector of the theory.
Our central result is to obtain the five--dimensional effective theory of
strongly coupled heterotic string for all $(1,1)$ fields and construct
its fundamental BPS domain wall three--brane solutions. We show that,
in the bulk, this theory is indeed a form of gauged supergravity. We
argue that this effective theory together with the three--brane
solutions is the proper starting point for particle phenomenology and
early universe cosmology. As an explicit demonstration we derive the
K\"ahler--potential, the superpotential and the gauge--kinetic
functions of the four--dimensional theory by reducing on the
five--dimensional theory on this three--brane solution. We also
comment on how gaugino condensation appears in this formalism. 

\vspace{0.4cm}

At first sight, it seems puzzling how a sensible five--dimensional theory
can be derived by reducing Ho\v rava--Witten theory. Generically,
the antisymmetric tensor field of the 11--dimensional theory has magnetic
sources provided by the $E_8$ gauge fields (and gravity) on the
orbifold hyperplanes. In particular, for compactification on a Calabi--Yau
three--fold using the standard embedding of the spin--connection into one
of the $E_8$ gauge groups one finds that the magnetic sources are nonzero
and of opposite strength. Hence, unlike in the weakly coupled heterotic
string, one is forced to consider backgrounds with a nonvanishing internal
antisymmetric tensor field strength. To satisfy the Killing spinor equations
in the presence of this field one is led to ``deform'' the Calabi--Yau
space and introduce dependence on the orbifold coordinates into the
background metric. This procedure gives the solutions of
ref.~\cite{w}, which were explicitly constructed in detail in
ref.~\cite{low1}. Although those solutions are perfectly appropriate
for a reduction to four dimensions~\cite{low1}, their explicit orbifold
dependence makes them unsuitable to derive a five--dimensional effective
action with the orbifold among those five dimensions.

Which backgrounds should then be used to derive the five--dimensional action?
An answer to this is provided by a change of perspective.
Rather than viewing the nonzero antisymmetric tensor field, which is the cause
of the metric deformation, as part of the pure background we will
keep this field as an ingredient of the five--dimensional theory.
As we will see, this internal antisymmetric tensor field is characterized
by a topologically nontrivial solution for the four--form field strength
on the Calabi--Yau space $X$ and can hence be identified with the
cohomology group $H^4(X)$. Such a configuration is called a nonzero mode
of the antisymmetric tensor field. Alternatively, the nonzero mode can be
described by a set of charges $\a_i$, $i=1,\cdots ,h^{1,1}$ associated
with the integrals of the boundary gauge--field and gravity sources over
the Calabi--Yau four--cycles. As those boundary fields form magnetic
sources for the antisymmetric tensor field, the charges $\a_i$ can be
interpreted as five--brane charges with the associated five--branes 
confined to the orbifold planes. In conclusion, the correct, consistent
reduction to five dimensions should then be performed on a conventional,
undeformed Calabi--Yau three--fold but keeping the nonzero mode for the
antisymmetric tensor field as dictated by the boundary sources.

The bulk action derived in such a way is a minimal $N=1$ supergravity
theory in five dimensions coupled to $h^{1,1}-1$ vector multiplets and the
universal hypermultiplet with sigma model coset $SU(2,1)/U(2)$. This action
is not identical with the one obtained by a ``conventional'' reduction of
pure 11--dimensional supergravity on a Calabi--Yau three--fold, as, for
example, carried out in ref.~\cite{CYred}. The difference is due to the
non--zero mode which leads to the gauging of a $U(1)$ subgroup in the
hypermultiplet isotropy group with the corresponding
gauge field being a certain linear combination of the vector multiplet
gauge fields and the gauge field in the gravity multiplet. In
particular it is the dual of the four-form antisymmetric field
strength in five-dimensions which is gauged. This can be understood as
a consequence of the Chern-Simons term in eleven dimensions. As a
result of this gauging, the theory contains a potential for
the Calabi--Yau breathing mode in the hypermultiplet and the Calabi--Yau
``shape'' moduli in the vector multiplets. The steepness of this
potential is set by the five--brane charges $\a_i$. Although
five--dimensional supergravity theories coupled to matter have been studied in
the literature~\cite{cn,GST1,GST2,Sierra} the general case with gauging
which we obtain from this reduction with non--zero mode has not been worked
out previously. We will close this gap in an appendix thereby establishing
that our result indeed represents a five--dimensional supergravity theory.
One notes that the appearance of gauged supergravity when non-zero
modes are included has been observed before in the context of
Calabi-Yau compactification of type II theories~\cite{sp,michelson}. 
In further contrast to the pure 11--dimensional supergravity case,
the bulk theory couples to four dimensional $N=1$ theories with gauge and
chiral multiplets confined to the now four--dimensional orbifold
hyperplanes. In addition, we find ``boundary potentials'' for the
projection of the bulk scalars onto the hyperplanes again specified by
the five--brane charges $\a_i$. Following ref.~\cite{hor,low2}, we will
also explain how to properly incorporate gaugino condensates into the
five--dimensional theory.

The existence of the potentials has important consequences for the
``vacuum structure'' of the five--dimensional theory. For a compact
Calabi--Yau space, the potential
terms are nonvanishing and, hence, flat space is not a solution of the
theory. Instead, we find the fundamental solution of the theory is a
new type of three--brane domain wall which couples to the bulk potential.
This is in analogy with the eight--brane in massive type IIA supergravity
which also couples to a scalar potential. The charges $\a_i$ play the
role of the mass in the eight--brane theory. More precisely, this
five--dimensional solution is a multi--charged double domain wall
since it carries all the charges $\a_i$ and has two sources located
one on each orbifold hyperplanes. Thus, each domain wall lies on one
of the two orbifold planes and carries the four--dimensional physical fields.
In this sense, upon reduction to four dimensions, $3+1$--dimensional
space--time is identified with the three--brane worldvolume.
We find that, although the equations of motion can be solved, the
three--brane can generally only be written implicitly in terms of the solution
of a system of quadratic equations depending on the Calabi--Yau intersection
numbers. In this form, however, the solution is quite general and applies
to any Calabi--Yau space (that is any number of moduli and any set of
intersection numbers). We also present various cases where the solution
can be made more explicit, one being a linearized approximation in the
charges $\a_i$. This linearized version of the three--brane coincides with
the zero mode part of the original deformed 11--dimensional
solution~\cite{w} which was also determined to leading order in the
charges. Hence, our three--brane represents a generalization of the
11--dimensional solution at all orders in the charges or, correspondingly,
to all orders in $\k^{2/3}$ in the five--dimensional action.

From the eleven--dimensional viewpoint the three--brane
originates from five--branes located on the 10--dimensional
orbifold planes, with two dimensions wrapped over a Calabi--Yau
two--cycle. It should be emphasized that this interpretation is not a
result of any approximation, but is an {\em exact} statement in the
context of the second-order effective theory that we consider here,
since the five-dimensional theory that we shall derive is obtained {\it via} a
consistent truncation of the eleven--dimensional theory. Consistency
of ordinary Kaluza-Klein reduction on Calabi-Yau manifolds to gravity
plus moduli fields was established in ref.\ \cite{dfps}. In the present
work and in \cite{losw}, such a consistent reduction is generalized to
include non-zero modes. The picture of wrapped 5-branes in $D=11$ also
provides an interpretation of the potential for the moduli which appears
in the five dimensional theory. Unlike the case for a conventional
Calabi--Yau compactification where all topologically equivalent
Calabi--Yau spaces are on the same footing (resulting in a flat moduli
space), in our case an expansion of the Calabi--Yau manifold reduces
the energy stored in the non--trivial antisymmetric tensor field. This
fact is reflected by the presence of the five--dimensional potential. 

The five--dimensional theory together with its three--brane
ground state is the correct starting point for low energy particle
phenomenology as well as early universe cosmology. In general, the low--energy
four--dimensional $N=1$ supergravity theory is obtained as a reduction of
this five--dimensional theory on the domain wall background. At the same
time, cosmologically interesting solutions are those which evolve into
the domain wall solution. Correspondingly, one should be interested in
cosmological solutions that depend on the orbifold coordinate as well as
on time~\cite{benakli,low4}. As an application, we will derive the
four--dimensional K\"ahler potential, the superpotential and the gauge
kinetic functions to linear order in the charges $\a_i$ for the full
$(1,1)$ sector of the theory. We find an $\a_i$ dependent correction
to the matter part of the K\"ahler potential and threshold corrections
to the gauge kinetic functions. These results are new in that they
cover all $(1,1)$ modes of the theory and they generalize the
universal expressions obtained in ref.~\cite{hp,low1}.

\vspace{0.4cm}

Let us now summarize our conventions. A more detailed account of the
conventions for Calabi--Yau spaces and five--dimensional supergravity is
given in the appendices. We will
consider eleven-dimensional spacetime compactified on a Calabi-Yau space $X$,
with the subsequent reduction down to four dimensions effectively provided by
a double-domain-wall background, corresponding to an $S^1/Z_2$ orbifold. We
use coordinates $x^{I}$ with indices $I,J,K,\ldots = 0,\ldots ,9,11$ to
parameterize the full 11--dimensional space $M_{11}$. Throughout this paper,
when we refer to orbifolds, we will work in the ``upstairs'' picture
with the orbifold $S^1/Z_2$ in the $x^{11}$--direction. We choose the range
$x^{11}\in [-\pi\rho ,\pi\rho ]$ with the endpoints being identified. The
$Z_2$ orbifold symmetry acts as $x^{11}\rightarrow -x^{11}$. There then exist
two ten--dimensional hyperplanes fixed under the $Z_2$ symmetry which we
denote by $M_{10}^{(n)}$, $n=1,2$. Locally, they are specified by the
conditions $x^{11}=0,\pi\rho$. Upon reduction on a Calabi--Yau space to
five dimensions they lead to four--dimensional fixed hyperplanes $M_4^{(n)}$.
Barred indices $\bar{I},\bar{J},\bar{K},\ldots = 0,\ldots ,9$ are used for the
ten--dimensional space orthogonal to the orbifold. 
Upon reduction on the Calabi-Yau space we have a five-dimensional spacetime
$M_5$ labeled by indices $\a ,\b ,\g ,\ldots  = 0,\ldots ,3,11$. The
orbifold fixed planes become four-dimensional with indices
$\m,\n,\rho,\ldots = 0,\ldots ,3$. We use indices $A,B,C,\ldots =
4,\ldots 9$ for the Calabi--Yau space. Holomorphic and anti--holomorphic
indices on the Calabi--Yau space are denoted by $a,b,c,\ldots$ and
$\bar{a},\bar{b},\bar{c},\ldots$, respectively. The harmonic $(1,1)$--forms of
the Calabi--Yau space on which we will concentrate throughout this paper
are indexed by $i,j,k,\ldots = 1,\ldots ,h^{1,1}$.

The 11-dimensional Dirac--matrices $\G^I$ with $\{\G^I,\G^J\}=2g^{IJ}$
are decomposed as $\G^I = \{\g^\a\otimes\l ,{\bf 1}\otimes\l^A\}$
where $\g^\a$ and $\l^A$ are the five-- and six--dimensional Dirac
matrices, respectively. Here, $\l$ is the chiral projection matrix in
six dimensions with $\l^2=1$. Spinors in eleven dimensions are
Majorana with 32 real components throughout the paper. In five
dimensions we use symplectic--real spinors~\cite{c0}. These are
defined in appendix B. 
Fields will be required to have a definite behavior under the $Z_2$
orbifold symmetry in $D=11$. We demand a bosonic field $\Phi$ to be
even or odd; that is, $\Phi (x^{11})=\pm\Phi (-x^{11})$. For a spinor
$\Psi$ the condition is $\G_{11}\Psi (-x^{11})=\pm\Psi (x^{11})$ and, depending
on the sign, we also call the spinor even or odd. The projection to one of
the orbifold planes leads then to a ten--dimensional Majorana--Weyl spinor
with definite chirality. Similarly, in five dimensions, bosonic fields will be
either even or odd, and there is a corresponding orbifold condition on
spinors. 


\section{Eleven--dimensional Supergravity on an Orbifold}


In this section we briefly review the formulation of the low--energy
effective action of strongly coupled heterotic string theory as
eleven--dimensional supergravity on the orbifold $S^1/Z_2$ due to
Ho\v{r}ava and Witten~\cite{hw1,hw2}. 

The bosonic part of the action is given by
\begin{equation}
\label{action}
   S = S_{\rm SG}+S_{\rm YM}
\end{equation}
where $S_{\rm SG}$ is the familiar 11--dimensional supergravity action 
\begin{equation}
 S_{\rm SG} = -\frac{1}{2\k^2}\int_{M^{11}}\sqrt{-g}\left[ 
                    R+\frac{1}{24}G_{IJKL}G^{IJKL}
           +\frac{\sqrt{2}}{1728}\e^{I_1...I_{11}}
               C_{I_1I_2I_3}G_{I_4...I_7}G_{I_8...I_{11}} \right]
 \label{SSG}
\end{equation}
and $S_{\rm YM}$ describes the two $E_8$ Yang--Mills theories on the
orbifold planes, explicitly given by~\footnote{We note that there is a
debate in the literature about the precise value of the Yang--Mills
coupling constant in terms of $\k$. While we quote the original
value~\cite{hw2,deA} the value found in ref.~\cite{conrad} is
smaller. In the second case, the coefficients in the Yang-Mills
action~\eqref{SYM} and the Bianchi identity~\eqref{Bianchi} should
both be multiplied by $2^{-1/3}$. This potential factor will not be
essential in the following discussion as it will simply lead to a
redefinition of the five--dimensional coupling constants. In the
following, we will give the necessary modifications where
appropriate.} 
\begin{multline}
   \label{SYM}
   S_{\rm YM} = - \frac{1}{8\pi\k^2}\left(\frac{\k}{4\pi}\right)^{2/3}
        \int_{M_{10}^{(1)}}\sqrt{-g}\;\left[
           \tr(F^{(1)})^2 - \frac{1}{2}\tr R^2\right] \\
        - \frac{1}{8\pi\k^2}\left(\frac{\k}{4\pi}\right)^{2/3}
           \int_{M_{10}^{(2)}}\sqrt{-g}\;\left[
               \tr(F^{(2)})^2 - \frac{1}{2}\tr R^2\right]\; .
\end{multline}
Here $F_{\bar{I}\bar{J}}^{(n)}$ are the two $E_8$ gauge field strengths and
$C_{IJK}$ is the 3--form with field strength
$G_{IJKL}=24\,\partial_{[I}C_{JKL]}$. The above action has to be supplemented
by the Bianchi identity
\begin{equation}
 (dG)_{11\bar{I}\bar{J}\bar{K}\bar{L}} = -\frac{1}{2\sqrt{2}\pi}
    \left(\frac{\k}{4\pi}\right)^{2/3} \left\{ 
       J^{(1)}\d (x^{11}) + J^{(2)}\d (x^{11}-\pi\r )
       \right\}_{\bar{I}\bar{J}\bar{K}\bar{L}} \label{Bianchi}
\end{equation}
where the sources are defined by 
\begin{equation}
 J^{(n)}
    = {\rm tr}F^{(n)}\wedge F^{(n)} 
      - \frac{1}{2}{\rm tr}R\wedge R \; .\label{J}
\end{equation}
Note that, in analogy with the weakly coupled case, the boundary $\tr R^2$
terms in eq.~\eqref{SYM} are required by supersymmetry as pointed out in
ref.~\cite{low1}. Under the $Z_2$ orbifold symmetry, the field components
$g_{\bar{I}\bar{J}}$, $g_{11,11}$, $C_{\bar{I}\bar{J}11}$ are even, while
$g_{\bar{I}11}$, $C_{\bar{I}\bar{J}\bar{K}}$ are odd. The above action
is complete to order $\k^{2/3}$ relative to the bulk. Corrections, however, 
will appear as higher--dimension operators at order $\k^{4/3}$.

The fermionic fields of the theory are the 11--dimensional gravitino
$\J_I$ and the two 10--dimensional Majorana--Weyl spinors $\c^{(n)}$, located
on the boundaries, one for each $E_8$ gauge group. The components
$\J_{\bar{I}}$ of the gravitino are even while $\J_{11}$ is odd. The gravitino
supersymmetry variation is given by
\begin{equation} 
 \d\J_I = D_I\eta +\frac{\sqrt{2}}{288}\left(\G_{IJKLM}-8g_{IJ}\G_{KLM}
          \right)G^{JKLM}\eta +\; \cdots \label{susy}\; ,
\end{equation}
where the dots indicate terms that involve fermion fields. The spinor $\eta$
in this variation is $Z_2$ even.

The appearance of the boundary source terms in the Bianchi identity
has a simple interpretation by analogy with the theory of
$D$-branes. It is well known that the $U(N)$ gauge fields describing
the theory of $N$ overlapping $Dp$-branes encode the charges for
lower-dimensional $D$-branes embedded in the $Dp$-branes. For
instance, the magnetic flux $\tr F$ couples to the $p-1$-form
Ramond-Ramond potential, so describes $D(p-2)$-brane charge. Higher
cohomology classes $\tr F\wedge \cdots \wedge F$ describe the
embedding of lower-dimensional branes. Furthermore, if the $Dp$-brane
is curved, then the cohomology classes of the tangent bundle also
contribute. For instance $\tr R\w R$ induces $D(p-4)$-brane charge. 
We recall that in eleven dimensions it is M five-branes which are
magnetic sources for $G_{IJKL}$. Thus we can interpret the magnetic
sources in the Bianchi identity~\eqref{Bianchi} as five-branes
embedded in the orbifold fixed planes. 

The relationship between the five-branes and the orbifold fixed planes
requires some comment. Above, the theory is formulated with explicit
delta-function sources describing the five-brane charges in the
orbifold planes. In the following sections, we will reduce the theory
on a Calabi-Yau, which means this five-brane charge is non-zero. In
the reduced theory the five-branes appear as three-brane domain
walls, localized in the fixed planes, which again provide the
necessary delta-function sources to support them. We will see that the
domain wall solution is the natural ``vacuum'' of the reduced
theory. However, we could also change perspective and view the theory
solely from the bulk eleven-dimensional, or reduced five-dimensional,
supergravity theory, which certainly have an independent existence
aside from the orbifold. Backgrounds with five-branes would then
appear as solitons carrying a magnetic charge. Similarly, the domain
wall solution is simply a soliton in the five-dimensional action. As
with all solitonic solutions to supergravity theories, however, static
vacuum solutions also have multiplets of zero-mode excitations
associated to them, and owing to the presence of worldvolume
anomalies, the spectrum of such zero-modes can be somewhat complex. It
would be interesting to investigate the structure of the effective
theory of the zero modes, analyzing the anomaly constraints along the
lines of the case of superconducting strings \cite{calhar} or M
five-branes \cite{fhmm}. One might expect that this would reproduce at
least some of the dynamics encoded in the orbifold plane Yang-Mills
source action given above, and the orbifold description could,
at least partially, be dispensed with. 

This speculation takes us outside the immediate context of the present
paper, however. For the meantime, we shall remain content to view the
five-dimensional solitonic domain walls as simply being embedded into
the orbifold fixed planes, whose curvature and instanton number
provide the necessary delta-function sources and which naturally then
lead to the five-brane charges associated to the domain walls in the
$D=11$ perspective. Physically, by analogy with the $D$-brane case, we
can identify the instanton number with the number of physical
five-branes living in the fixed-planes, while the curvature leads to
an induced five-brane charge. This interpretation will prove a useful
guide in understanding the structure of the reduced theory and its
vacuum solution in the compactification to five dimensions.


\section{The five--dimensional effective theory}


As mentioned in the introduction, matching of scales suggests that strongly
coupled heterotic string theory appears effectively five--dimensional in
some intermediate energy range. In this section we derive the
five--dimensional effective theory in this regime obtained by a
compactification on a Calabi--Yau three--fold. We expect that this
should lead to a theory with bulk $N=1$ five-dimensional supersymmetry
and four-dimensional $N=1$ supersymmetry on the orbifold fixed
planes. As we will see, doing this compactification 
consistently requires the inclusion of non--zero modes for the field strength
of the anti--symmetric tensor field. These non--zero modes appear in the
purely internal Calabi--Yau part of the anti--symmetric tensor field and
correspond to harmonic $(1,1)$ forms on the Calabi--Yau three--fold.
Consequently, to capture the complete structure of these non--zero modes, we
will have to consider the full $(1,1)$ sector of the theory. We will not, 
however, explicitly include the $(2,1)$ sector as it is largely unaffected by
the specific structure of Ho\v{r}ava--Witten theory. Instead, we comment on
the additions necessary to incorporate this sector along the way.

To make contact with the compactifications to four-dimensions
discussed by Witten~\cite{w}, it is very natural to embed the
spin-connection of the Calabi--Yau manifold in the gauge connection of
one of the $E_8$ groups breaking it to $E_6$. In general, this implies
that there is a non-zero instanton number on one of the orbifold planes.
From the discussion of the previous section, this can be interpreted as
including five-branes living in the orbifold plane in the
compactification. It is this additional element to the
compactification which introduces the non--zero mode and leads to much
of the interesting structure of the five-dimensional theory. We note
that the presence of five-brane charge is really unavoidable. Even
without exciting instanton number, the curvature of the Calabi-Yau
leads to an induced magnetic charge in the Bianchi
identity~\eqref{Bianchi}, forcing us to include non-zero modes.

\subsection{Zero modes}

Let us now explain the structure of the zero mode fields used in the reduction
to five dimensions. We begin with the bulk. The background space--time
manifold is $M_{11}=X\times S_1/Z_2\times M_4$, where $X$ is
a Calabi--Yau three--fold and $M_4$ is four--dimensional Minkowski space.
Reduction on such a background leads to eight preserved supercharges and,
hence, to minimal $N=1$ supergravity in five dimensions. Due to the
projection condition, this leads to four preserved supercharges on the
orbifold planes implying four--dimensional $N=1$ supersymmetry on those
planes. Including the zero modes, the metric is given by
\begin{equation}
 ds^2 = V^{-2/3}g_{\a\b}dx^\a dx^\b +g_{AB}dx^Adx^B \label{metric}
\end{equation}
where $g_{AB}$ is the metric of the Calabi--Yau space $X$. Its K\"ahler form
is defined by~\footnote{Note here that we choose the opposite sign convention
as in ref.~\cite{w,low1} to conform with the literature on Calabi--Yau
reduction of 11--dimensional supergravity and type II
theories~\cite{CYred,bcf}.} $\o_{a\bar{b}}=ig_{a\bar{b}}$ and can be expanded
in terms of the harmonic $(1,1)$--forms $\o_{iAB}$, $i=1,\cdots ,h^{1,1}$ as
\begin{equation}
 \o_{AB} = a^i\o_{iAB}\; .\label{ai_def}
\end{equation}
The coefficients $a^i=a^i(x^\a )$ are the $(1,1)$ moduli of the Calabi--Yau
space. The Calabi--Yau volume modulus $V=V(x^\a )$ is defined by
\begin{equation}
 V=\frac{1}{v}\int_X\sqrt{^6g}\label{V_def}
\end{equation}
where $^6g$ is the determinant of the Calabi--Yau metric $g_{AB}$. In
order to make $V$ dimensionless we have introduced a coordinate volume
$v$ in this definition which can be chosen for convenience. The modulus
$V$ then measures the Calabi--Yau volume in units of $v$. The factor
$V^{-2/3}$ in eq.~\eqref{metric} has been chosen such that the metric
$g_{\a\b}$ is the five--dimensional Einstein frame metric. Clearly $V$
is not independent of the $(1,1)$ moduli $a^i$ but it can be expressed as
\begin{equation}
 V = \frac{1}{6}\cK (a)\; ,\quad \cK (a) = d_{ijk}a^ia^ja^k
\end{equation}
where $\cK (a)$ is the K\"ahler potential and $d_{ijk}$ are the Calabi--Yau
intersection numbers. Their definition, along with a more detailed account
of Calabi--Yau geometry, can be found in appendix A.

Let us now turn to the zero modes of the antisymmetric tensor field.
We have the potentials and field strengths,
\bea
 C_{\a\b\g}&,\quad& G_{\a\b\g\d}\nn \\
 C_{\a AB}=\frac{1}{6}\cA_\a^i\o_{iAB}
           &,\quad&G_{\a\b AB}=\cF_{\a\b}^i\o_{iAB} \\
 C_{abc} =\frac{1}{6}\x\O_{abc}&,\quad&G_{\a abc}=X_\a\O_{abc}\nn \; .
 \label{Czero}
\eea
The five--dimensional fields are therefore an antisymmetric
tensor field $C_{\a\b\g}$ with field strength $G_{\a\b\g\d}$, $h^{1,1}$
vector fields $\cA_\a^i$ with field strengths $\cF_{\a\b}^i$ and a
complex scalar $\x$ with field strength $X_\a$ that arises from the harmonic
$(3,0)$ form denoted by $\O_{abc}$. In the bulk the relations
between those fields and their field strengths are simply
\bea
 G_{\a\b\g\d} &=& 24\,\partial_{[\a}C_{\b\g\d ]} \nn \\
 \cF_{\a\b}^i &=& \partial_\a\cA_\b^i-\partial_\b\cA_\a^i \\
 X_\a &=& \partial_\a\x\nn \; .
\eea
These relations, however, will receive corrections from the boundary
controlled by the 11--dimensional Bianchi identity~\eqref{Bianchi}. We will
derive the associated five--dimensional Bianchi identities later.

\vspace{0.4cm}

Next, we should set up the structure of the boundary fields. The starting
point is the standard embedding of the spin connection in
the first $E_8$ gauge group such that
\begin{equation}
 \tr F^{(1)}\w F^{(1)} = \tr R\w R\; .
\label{condition}
\end{equation}
As a result, we have an $E_6$ gauge field $A_\a^{(1)}$ with field strength
$F_{\m\n}^{(1)}$ on the first hyperplane and an $E_8$ gauge field $A_\m^{(2)}$
with field strength $F_{\m\n}^{(2)}$ on the second hyperplane. In addition,
there are $h^{1,1}$ gauge matter fields from the $(1,1)$ sector on the first
plane. They are specified by
\begin{equation}
 A_b^{(1)} = \bar{A}_b+{\o_{ib}}^cT_{cp}C^{ip}
\end{equation}
where $\bar{A}_b$ is the (embedded) spin connection. Furthermore,
$p,q,r,\ldots =1,\ldots ,27$ are indices in the fundamental ${\bf 27}$
representation of $E_6$ and $T_{ap}$ are the $({\bf 3},{\bf 27})$
generators of $E_8$ that arise in the decomposition under the subgroup
$SU(3)\times E_6$. Their complex conjugate is denoted by
$T^{ap}$. The $C^{ip}$ are $h^{1,1}$ complex scalars in the
${\bf 27}$ representation of $E_6$. Useful traces for these generators
are $\tr (T_{ap}T^{bq})=\d_a^b\d_p^q$ and
$\tr (T_{ap}T_{bq}T_{cr})=\O_{abc}f_{pqr}$ where $f_{pqr}$ is the totally
symmetric tensor that projects out the singlet in ${\bf 27}^3$.

\subsection{The nonzero mode}

So far, what we have considered is similar to a reduction of pure
11--dimensional supergravity on a Calabi--Yau space, as for example
performed in ref.~\cite{CYred}, with the addition of gauge and gauge
matter fields on the boundaries. An important difference arises, however,
because the standard embedding~\eqref{condition}, unlike in the case
of the weakly coupled heterotic string, no longer leads to vanishing
sources in the Bianchi identity~\eqref{Bianchi}. Instead, there is a
net five-brane charge, with opposite sources on each fixed plane,
proportional to $\pm\tr R\w R$. The nontrivial components of
the Bianchi identity~\eqref{Bianchi} are given by 
\begin{equation}
 (dG)_{11ABCD} = -\frac{1}{4\sqrt{2}\pi}
    \left(\frac{\k}{4\pi}\right)^{2/3} \left\{ 
       \d (x^{11}) - \d (x^{11}-\pi\r )\right\} (\tr R\w R)_{ABCD}\; .
 \label{Bianchi1}
\end{equation}
As a result, the components $G_{ABCD}$ and $G_{ABC11}$ of the antisymmetric
tensor field are nonvanishing. More precisely, the above equation has to be
solved along with the equation of motion. 
\begin{equation}
 D_IG^{IJKL} = 0\; . \label{Geom}
\end{equation}
(Note that the Chern--Simons contribution to the antisymmetric tensor
field equation of motion vanishes if $G_{ABCD}$ and $G_{ABC11}$ are
the only nonzero components of $G_{IJKL}$.) 
The general solution of these equations is quite complicated and has
been given in ref.~\cite{low1} as an expansion in Calabi--Yau harmonic
functions. For the present purpose of deriving a 
five--dimensional effective action, we are only interested in the zero
mode terms in this expansion because the heavy Calabi--Yau modes decouple
as a result of the consistent Kaluza-Klein truncation to $D=5$. 
To work out the zero mode part of the solution, we note
that $\tr R\w R$ is a $(2,2)$ form on the Calabi--Yau space (since the
only nonvanishing components of a Calabi--Yau curvature tensor are
$R_{a\bar{b}c\bar{d}}$). Let us, therefore, introduce a basis $\n^i$,
$i=1,\cdots ,h^{2,2}=h^{1,1}$ of harmonic $(2,2)$ forms and
corresponding four--cycles $C_i$ such that
\begin{equation}
 \frac{1}{v}\int_X \o_i\w \n^j = \d_i^j\; ,\quad 
  \frac{1}{v^{2/3}}\int_{C_i}\n^j = \d_i^j\; .
\end{equation}
The zero mode part $\left.\tr R\w R\right|_0$ of the source can then be
expanded as
\begin{equation}
 \left.\tr R\w R\right|_0 = -8\sqrt{2}\p \left(\frac{4\p}{\k}
                             \right)^{2/3}\, \a_i\,\n^i \label{expans}
\end{equation}
where the numerical factor has been included for convenience. The
expansion coefficients $\a_i$ are
\begin{equation}
 \a_i =
 \frac{\pi}{\sqrt{2}}\left(\frac{\k}{4\p}\right)^{2/3}\frac{1}{v^{2/3}}
 \b_i\; ,\qquad \b_i=-\frac{1}{8\pi^2}\int_{C_i}\tr R\w R \; .
 \label{alpha_def}
\end{equation}
Note that $\b_i$ are integers, characterizing the first Pontrjagin
class of the Calabi-Yau. It is then straightforward to see that the
zero mode part of the Bianchi identity~\eqref{Bianchi1} and the
equation of motion~\eqref{Geom} are 
solved by
\bea
 \left. G_{ABCD}\right|_0 &=& \a_i\,\n^i_{ABCD}\,\e (x^{11})=\frac{1}{4V}\a^i\,
                 {\e_{ABCD}}^{EF}\,\o_{iEF}\,\e (x^{11})\label{nonzero} \\
 \left.G_{ABC11}\right|_0 &=& 0\; .
\eea
Here $\e (x^{11})$ is the step function which is $+1$ for positive
$x^{11}$ and $-1$ otherwise. The index of the coefficient $\a^i$ in the
second part of the first equation has been raised using the metric
\begin{equation}
 G_{ij}(a) = \frac{1}{2V}\int_X\o_i\w (*\o_j) \label{CY_metric}
\end{equation}
on the $(1,1)$ moduli space. Note that, while the coefficients $\a_i$ with
lowered index are truly constants, as is apparent from
eq.~\eqref{alpha_def}, the coefficients $\a^i$ depend on the $(1,1)$
moduli $a^i$ since the metric~\eqref{CY_metric} does. From the
expansion~\eqref{expans} we can derive an expression for the boundary
$\tr F^2$ and $\tr R^2$ terms in the action~\eqref{SYM} which will be
essential for the reduction of the boundary theories. We have
\begin{equation} 
 \left.\tr R_{AB}R^{AB}\right|_0 = \left.\tr F^{(1)}_{AB}F^{(1)AB}\right|_0 =
  -4\sqrt{2}\p\left(\frac{4\p}{\k}\right)^{2/3}V^{-1}\a^i\o^{AB}\o_{iAB}
 \label{R2}
\end{equation}
while, of course
\begin{equation}
 \tr F^{(2)}_{AB}F^{(2)AB} = 0\; .
\end{equation}

The expression~\eqref{nonzero} for $G_{ABCD}$ with $\a_i$ as defined in
\eqref{alpha_def} is the new and somewhat unconventional ingredient in our
reduction. Using the terminology of ref.~\cite{gsw} we call this configuration
for the antisymmetric tensor field strength a nonzero mode. Generally,
a nonzero mode is defined as a nonzero internal antisymmetric tensor
field strength $G$ that solves the equation of motion. In contrast,
conventional zero modes of an antisymmetric tensor field, like those
in eq.~\eqref{Czero}, have vanishing field strength once the moduli fields
are set to constants. Since the kinetic term $G^2$ is positive for a nonzero
mode it corresponds to a nonzero energy configuration. Given that
nonzero modes, for a $p$--form field strength, satisfy
\begin{equation}
 dG=d^*G=0
\end{equation}
they correspond to harmonic forms of degree $p$. Hence, they can
be identified with the $p$th cohomology group $H^p(X)$ of the internal
manifold $X$. In the present case, we are dealing with a four--form
field strength on a Calabi--Yau three--fold $X$ so that the relevant
cohomology group is $H^4(X)$. The expression~\eqref{nonzero} is just an
expansion of the nonzero mode in terms of the basis $\{\n^i\}$ of
$H^4(X)$. The appearance of all harmonic $(2,2)$ forms shows that it is
necessary to include the complete $(1,1)$ sector into the low energy effective
action in order to fully describe the nonzero mode, as argued in the
beginning of this section. On the other hand, harmonic $(2,1)$ forms do not
appear here and are hence less important in our context. We stress that
the nonzero mode~\eqref{nonzero}, for a given Calabi--Yau space, specifies
a fixed element in $H^4(X)$ since the coefficients $\a_i$ are fixed in terms
of Calabi--Yau properties. In fact, they are related to the integers
$\b_i$ characterizing the first Pontrjagin class of the tangent
bundle. Thus we see that, correctly normalized, $G$ is in the integer
cohomology of the Calabi-Yau. This quantization condition has been
described in~\cite{wittq}

In a dimensional reduction of pure 11--dimensional supergravity, non--zero
modes can be considered as well but are usually dismissed as non--zero energy
configuration. Compactifications of 11--dimensional supergravity on various
manifolds including Calabi--Yau three-folds with non--zero modes have
been considered in the literature~\cite{llp}. The difference in our case is
that we are not free to turn off the non--zero mode. Its presence is
simply dictated by the nonvanishing boundary sources.

\subsection{The five--dimensional effective action}

Let us now summarize the field content which we have obtained above and
discuss how it fits into the multiplets of five--dimensional $N=1$
supergravity. The form of these multiplets and in particular the
conditions on the fermions is discussed in more detail
in appendix B. We know that the gravitational multiplet should contain
one vector field, the graviphoton. Thus since the reduction leads to
$h^{1,1}$ vectors, we must have $h^{1,1}-1$ vector multiplets. This
leaves us with the $h^{1,1}$ scalars $a^i$, the complex scalar $\xi$
and the three-form $C_{\a\b\g}$. Since there is one scalar in each
vector multiplet, we are left with three unaccounted for real scalars
(one from the set of $a^i$, and $\xi$) and the three-form. Together,
these fields form the ``universal hypermultiplet;'' universal because
it is present independently of the particular form of the
Calabi-Yau manifold. From this, it is clear that it must be the overall
volume breathing mode  $V=\frac{1}{6}d_{ijk}a^ia^ja^k$ that is the
additional scalar from the set of the $a^i$ which enters the universal
multiplet. The three-form may appear a little unusual, but one
should recall that in five dimensions a three-form is dual to a scalar
$\s$. Thus, the bosonic sector of the universal hypermultiplet consists
of the four scalars $(V,\s,\xi,\bar{\xi})$. 

The $h^{1,1}-1$ vector multiplet scalars are the remaining $a^i$. More
properly, since the breathing mode $V$ is already part of a
hypermultiplet it should be first scaled out when defining the shape
moduli 
\begin{equation}
 b^i=V^{-1/3}a^i\; .\label{bi_def}
\end{equation}
Note that the $h^{1,1}$ moduli $b^i$ represent only $h^{1,1}-1$
independent degrees of freedom as they satisfy the constraint
\begin{equation}
 \cK (b)\equiv d_{ijk}b^ib^jb^k =6\; .
\end{equation}
Alternatively, as described in appendix B, we can introduce
$h^{1,1}-1$ independent fields $\phi^x$ with $b^i=b^i(\phi^x)$. The
bosonic fields in the vector multiplets are then given by
$(\phi^x,b^x_i\cA^i_\a)$ ($b^x_i$ represents a projection onto the
$\phi^x$ subspace). Meanwhile the graviton and graviphoton of the
gravity multiplet are given by
$(g_{\a\b},\frac{2}{3}b_i\cA^i_\a)$. Again, the details of this
decomposition are described in appendix B. 

Therefore, in total, the five dimensional bulk theory contains a gravity
multiplet, the universal hypermultiplet and $h^{1,1}-1$ vector multiplets.
The inclusion of the $(2,1)$ sector of the Calabi--Yau space would lead
to an additional $h^{2,1}$ set of hypermultiplets in the theory. Since
they will not play a prominent r\^ole in our context they will not be
explicitly included in the following.

On the boundary $M^{(1)}_4$ we have an $E_6$ gauge multiplet
$(A_\m^{(1)},\c^{(1)})$ and $h^{1,1}$ chiral multiplets $(C^{ip},\eta^{ip})$
in the fundamental ${\bf 27}$ representation of $E_6$. Here $C^{ip}$ denote
the complex scalars and $\eta^{ip}$ the chiral fermions. The other boundary,
$M^{(2)}_4$, carries an $E_8$ gauge multiplet $(A_\m^{(2)},\c^{(2)})$ only.
Inclusion of the $(2,1)$ sector would add $h^{2,1}$ chiral multiplets in
the $\bf{\overline{\bf 27}}$ representation of $E_6$ to
the field content of the boundary $M_4^{(1)}$. Any even bulk field
will also survive on the boundary. Thus, in addition to the
four--dimensional part of the metric, the scalars $b^i$ together with
$\cA^i_{11}$, and $V$ and $\s$ survive on the boundaries. These pair
into $h^{1,1}$ chiral muliplets. 

\vspace{0.4cm}

After this survey we are ready to derive the bosonic part of the
five--dimensional effective action for the $(1,1)$ sector. Inserting the
expressions for the various fields from the previous subsection
into the action~\eqref{action}, using the formulae given in appendix A and
dropping higher derivative terms we find
\begin{equation}
 S_5 = S_{\rm grav,vec}+S_{\rm hyper}+S_{\rm bound}+S_{\rm matter}
 \label{S5}
\end{equation}
with
\bsea
 S_{\rm grav,vec} &=& -\frac{1}{2\k_5^2}\int_{M_5}\sqrt{-g}\left[R+
                      G_{ij}\pt_\a b^i \pt^\a b^j +
                      \right.\nn \\
                   && \qquad\qquad\qquad\qquad \left.
                     G_{ij}\cF_{\a\b}^i\cF^{j\a\b}+\frac{\sqrt{2}}{12}
                      \e^{\a\b\g\d\e}d_{ijk}\cA_\a^i\cF_{\b\g}^j\cF_{\d\e}^k
                      \right]\\
 S_{\rm hyper} &=& -\frac{1}{2\k_5^2}\int_{M_5}\sqrt{-g}\left[
                   \frac{1}{2}V^{-2}\partial_\a V\partial^\a V
                   +2V^{-1}X_\a\bar{X}^\a
                   +\frac{1}{24}V^2G_{\a\b\g\d}G^{\a\b\g\d}
                   \right.\nn \\
                && \qquad\qquad\qquad\qquad
                    +\frac{\sqrt{2}}{24}\e^{\a\b\g\d\e}G_{\a\b\g\d}
                   \left(i(\x\bar{X}_\e-\bar{\x}X_\e )+
                   2\e (x^{11})\a_i\cA_\e^i\right)\nn\\
                &&\left.\qquad\qquad\qquad\qquad
                  +\frac{1}{2}V^{-2}G^{ij}\a_i\a_j\right]\qquad\\
 S_{\rm bound} &=& \frac{\sqrt{2}}{\k_5^2}\int_{M_4^{(1)}}\sqrt{-g}
                   \, V^{-1}\a_ib^i
                   -\frac{\sqrt{2}}{\k_5^2}\int_{M_4^{(2)}}\sqrt{-g}\,
                   V^{-1}\a_ib^i \\
 S_{\rm matter} &=& -\frac{1}{16\p\a_{\rm GUT}}
                   \sum_{n=1}^2\int_{M_4^{(n)}}\sqrt{-g}\, V\tr
                   {F_{\m\n}^{(n)}}^2\nn \\
                 && -\frac{1}{2\p\a_{\rm GUT}}
                    \int_{M_4^{(1)}}\sqrt{-g}\left[ G_{ij}(D_\m C)^i
                    (D^\m\bar{C})^j\right.\nn\\
                 &&\left.\qquad\qquad\qquad\qquad\qquad\qquad
                    +V^{-1}G^{ij}\frac{\partial W}
                    {\partial C^{ip}}\frac{\partial\bar{W}}
                    {\partial\bar{C}^j_p}+D^{(u)}D^{(u)}\right] \; .
                    \label{actparts}
\esea
All fields in this action that originate from the 11--dimensional
antisymmetric tensor field are subject to a nontrivial Bianchi identity.
Specifically, from eq.~\eqref{Bianchi} we have
\bsea
 (dG)_{11\m\n\r\s} &=& -\frac{\k_5^2}{4\sqrt{2}\pi\a_{\rm GUT}}\left\{ 
       J^{(1)} \d (x^{11})+ J^{(2)} \d (x^{11}-\pi\r )
       \right\}_{\m\n\r\s} \\
 (d\cF^i)_{11\m\n} &=& -\frac{\k_5^2}{4\sqrt{2}\pi\a_{\rm GUT}}J_{\m\n}^i
                       \d (x^{11}) \\
 (dX)_{11\m} &=& -\frac{\k_5^2}{4\sqrt{2}\pi\a_{\rm GUT}}J_\m
                 \d (x^{11}) \label{Bianchi5}
\esea
with the currents defined by
\bsea
 J_{\m\n\r\s}^{(n)}  &=& \left(\tr F^{(n)}\w F^{(n)}-\frac{1}{2}\tr R\w R
                         \right)_{\m\n\r\s} \\
 J_{\m\n}^i &=& -2iV^{-1}\G^i_{jk}\left((D_\m C)^{jp}(D_\n\bar{C})^k_p
                -(D_\m\bar{C})^k_p(D_\n C)^{jp}\right) \\
 J_\m &=& -\frac{i}{2}V^{-1}d_{ijk}f_{pqr}(D_\m C)^{ip}C^{jq}C^{kr}\; .
  \label{currents5}
\esea
The five--dimensional Newton constant $\k_5$ and the
Yang--Mills coupling $\a_{\rm GUT}$ are expressed in terms of
11--dimensional quantities as~\footnote{These relations are given
for the normalization of the 11--dimensional action as in eq.~\eqref{action}.
If instead the normalization of~\cite{conrad} is used the expression for 
$\a_{\rm GUT}$ gets rescaled to 
$a_{\rm GUT}=2^{1/3}\left(\k^2/2v\right)\left(4\p/\k\right)^{2/3}$ 
Otherwise the action and Bianchi identities are unchanged, except that in 
the expression~\eqref{alpha_def} for $\alpha_i$ the RHS is multiplied 
by $2^{1/3}$.}
\begin{equation}
 \k_5^2=\frac{\k^2}{v}\; ,\qquad \a_{\rm GUT} = \frac{\k^2}{2v}\left(
   \frac{4\p}{\k}\right)^{2/3}\; . \label{fdconst}
\end{equation}
We still need to define various quantities in the above action. The
metric $G_{ij}$ is given in terms of the K\"ahler potential $\cK$ as
\begin{equation}
 G_{ij}=-\frac{1}{2}\frac{\partial}{\partial b^i}\frac{\partial}
           {\partial b^j}\ln \cK\; .
\end{equation}
The corresponding connection $\G_{jk}^i$ is defined as
\begin{equation}
 \G_{jk}^i=\frac{1}{2}G^{il}\frac{\partial G_{jk}}{\partial b^l}\; .
\end{equation}
We recall that
\begin{equation}
 \cK = d_{ijk}b^ib^jb^k\; ,\label{kpot}
\end{equation}
where $d_{ijk}$ are the Calabi--Yau intersection numbers. All indices
$i,j,k,\cdots$ in the five--dimensional theory are raised and lowered with
the metric $G_{ij}$. A more explicit form of this metric can be found in
appendix A. We also recall that the fields $b^i$ are subject to the constraint
\begin{equation}
 \cK = 6 \label{b_cons}
\end{equation}
which should be taken into account when equations of motion are derived
from the above action. Most conveniently, it can be implemented by adding
a Lagrange multiplier term $\sqrt{-g}\l(\cK (b)-6)$ to the bulk action.
Furthermore, we need to define the superpotential
\begin{equation}
 W=\frac{1}{6}d_{ijk}f_{pqr}C^{ip}C^{jq}C^{kr}
\label{superpot}
\end{equation}
and the D--term
\begin{equation}
 D^{(u)} =G_{ij}\bar{C}^jT^{(u)}C^i
\label{Dterm}
\end{equation}
where $T^{(u)}$, $u=1,\ldots ,78$ are the $E_6$ generators in the fundamental
representation. The consistency of the above theory has been explicitly
checked by a reduction of the 11--dimensional equations of motion.

\vspace{0.4cm}

The most notable features of this action, at first sight, are the bulk
and boundary potentials for the $(1,1)$ moduli $V$ and $b^i$ that
appear in $S_{\rm hyper}$ and $S_{\rm bound}$. Those potentials involve
the five--brane charges $\a_i$, defined by eq.~\eqref{alpha_def}, that
characterize the nonzero mode. The bulk potential in the hypermultiplet
part of the action arises directly from the kinetic term $G^2$ of the
antisymmetric tensor field with the expression~\eqref{nonzero} for the
nonzero mode inserted. It can therefore be interpreted as the energy
contribution of the nonzero mode. The origin of the boundary potentials,
on the other hand, can be directly seen from eq.~\eqref{R2} and
the boundary actions~\eqref{SYM}. Essentially, they arise because the
standard embedding leads to nonvanishing internal boundary actions due
to the crucial factor $1/2$ in front of the $\tr R^2$ terms. This is in
complete analogy with the appearance of nonvanishing sources in the
internal part of the Bianchi identity which led us to introduce the
nonzero mode. 

Before we discuss the relation to five--dimensional supergravity theories
we would like to explain how gaugino condensation can be incorporated
into the above action. In ref.~\cite{hor} it was shown that gaugino
bilinears in the 11--dimensional action can be grouped into a perfect
square together with the antisymmetric tensor field kinetic term. 
Furthermore, it was pointed out in that paper, that the problem
arising from the term quartic in the gauginos which is proportional
to $\left[\d (x^{11})\right]^2$, can be resolved by a redefinition of
the antisymmetric tensor field that absorbs the condensate. The coupling
of the condensate and the antisymmetric tensor is then shifted to the
11--dimensional Bianchi identity. Based on this observation, a smooth
background with a gaugino condensate has been given in~\cite{hor} and
this was then used in ref.~\cite{low2} to derive a well defined
four--dimensional nonperturbative superpotential. In the
five--dimensional setting, a covariantly constant condensate
proportional to the holomorphic three--form
$\O_{abc}$ on the Calabi--Yau manifold groups itself into a perfect
square together with the corresponding zero mode; that is, the complex
field $\x$ in the hypermultiplet. Clearly, this perfect square
also involves a term proportional to $\left[\d (x^{11})\right]^2$ and
therefore leads to a similar problem to that in 11 dimensions. Of course,
we can apply the same trick and redefine the field $\x$. Then the
condensate disappears from the five--dimensional action but leads to
additional source terms in the Bianchi identity~\eqref{Bianchi5} for
$X_\a$. Having done this, one can expect to find smooth solutions of the
five--dimensional theory in the presence of the condensate. More
precisely, one should find a domain wall solution similar to the one we
will present in section 5.


\section{Relation to five-dimensional supergravity theories}


As we have argued in the previous section, the five-dimensional effective
action~\eqref{S5} should have $N=1$, $D=5$ supersymmetry in the bulk
and $N=1$, $D=4$ supersymmetry on the boundary. In this section, we will
rewrite the action in a supersymmetric form. This will allow us to
complete the action~\eqref{actparts} to include fermionic terms and
give the supersymmetry transformations. One thing we will not do is
complete the supersymmetry transformations to include the bulk and
boundary couplings, but we assume a consistent completion is possible,
as in eleven dimensions. 

Of particular interest is the presence of potential terms
in the bulk theory. Such terms are forbidden unless the theory is
gauged; that is, unless some of the fields are charged under Abelian
gauge fields $\cA_\a$. In order to identify the supersymmetry
structure of the theory in hand, we derive, in Appendix B, the general
form of gauged $D=5$ $N=1$ with charged hypermultiplets, borrowing
heavily from the work of G\"unaydin \textit{et al.}~\cite{GST1,GST2}
and Sierra~\cite{Sierra}, and from the general theory of gauged $D=4$,
$N=2$ supergravity as given, for instance, in~\cite{andetal}. 

Let us start by giving the $N=1$ structure of the four-dimensional
boundary theory. As discussed above, we have a set of chiral multiplets
with scalar components $C^{ip}$, together with vector multiplets with
gauge fields $A_\m^{(i)}$. (The vectors live on both boundaries, but
the chiral matter lives only on the $E_6$ boundary.) In addition, the
scalars from the bulk $(A,\s)$ and $(b^i,\cA^i_{11})$ also form chiral
multiplets. From the form of the theory on the boundaries we can give
explicitly the functions determining the $N=1$ theory. We have already
given the form of the superpotential and the $D$-term on the $E_6$
boundary in equations~\eqref{superpot} and~\eqref{Dterm}. It is also easy
to read off the K\"ahler potential on the $E_6$ boundary and the
gauge kinetic functions on either fixed plane. We find, without care
to correct normalizations,
\begin{equation}
   K = G_{ij}C^{ip}{\bar C}^i_p \qquad 
   f^{(n)} = V + i\s
\label{Kfs}
\end{equation}
The appearance of $\s$ in the gauge kinetic function is not
immediately apparent from the action~\eqref{S5}. However, it is easy
to show that on making the dualization of $C_{\a\b\g}$ to $\s$, which
is described in more detail below, the magnetic source in the Bianchi
indentity~\eqref{Bianchi5} for $C_{\a\b\g}$, becomes an electric
source for $\s$. The result is that the gauge kinetic terms in the
boundary action are modified to 
\begin{equation}
  -\frac{1}{16\p\a_{\rm GUT}} \sum_{n=1}^2 
        \int_{M_4^{(n)}}\sqrt{-g}\, \left[ 
          V \tr F_{\m\n}^{(n)}F^{(n)\m\n} 
          - \frac{\s}{2}\e^{\m\n\r\s}\tr F_{\m\n}^{(n)}\tr
             F_{\r\s}^{(n)} \right]
\end{equation}
One notes that the expressions~\eqref{Kfs} include dependence on the
bulk fields $b^i$ and $V$, evaluated on the appropriate
boundary. Further, we are considering the bulk multiplets as
parameters, as their dynamics comes from bulk kinetic terms. 

Now let us turn to the bulk theory. Our goal will be to identify the
action~\eqref{actparts} with the bosonic part of the general gauged
theory discussed in appendix B. The gauged theory is
characterized by a special Riemannian manifold $\cM_V$ describing the
vector multiplet sigma-model, a quaternionic manifold $\cM_H$
describing the hypermultiplet sigma-model, and a set of Killing vectors
and prepotentials on $\cM_H$. These are the structures we must
identify in the action~\eqref{actparts}.

We start by concentrating on the hypermultiplet structure. We have
argued that, after dualizing the three-form potential $C_{\a\b\g}$ to a
scalar $\s$, the fields $(V,\s,\xi,\xib)$ represent the scalar components of a
hypermultiplet. Concentrating on the kinetic terms let us make the
dualization explicit. We find
\begin{equation}
   G_{\a\b\g\d} = \frac{1}{\sqrt{2}}V^{-2}{\e_{\a\b\g\d}}^\e\left\{
      \pt_\e\s - i\left( \x\pt_\e\xib -\xib\pt_\e\x \right)
      - 2\e (x^{11})\a_i\cA^i_\e \right\} \; .
\end{equation}
The kinetic terms can then be written in the form
\begin{equation}
   h_{uv} D_\a q^u D^\a q^v 
\label{Hke}
\end{equation}
where $q^u=(V,\s,\x,\xib)^u$ and 
\begin{equation}
   D_\a q^u = \left(\pt_\a V,\, \pt_\a\s - 2\e (x^{11})\a_i\cA^i_\a,\, 
                  \pt_\a\x,\, \pt_\a\xib \right)^u
\label{covDer}
\end{equation}
and the metric is given by
\begin{equation}
   h_{uv} dq^u dq^v = \frac{1}{4V^2} dV^2 
       + \frac{1}{4V^2} \left[d\s+i(\x d\xib-\xib d\xi)\right]^2 
       + \frac{1}{V} d\xi d\xib \; .
\end{equation}
This reproduces the well-known result that the universal multiplet
classically parameterizes the quaternionic space
$\cM_H=SU(2,1)/U(2)$~\cite{quat}. 

In what follows, we would like to have an explicit realization of the
quaternionic structure of $\cM_H$. A review of quaternionic geometry
is given in Appendix B. We will now give expressions for the
quantities defined there, following a discussion given
in~\cite{strom}. Since we have a single hypermultiplet, the holonomy
of $\cM_H$ should be $SU(2)\times Sp(2)=SU(2)\times SU(2)$. To
distinguish these, we will refer to the first factor as
$SU(2)$ and the second as $Sp(2)$. Defining the symplectic matrix
$\O_{ab}$ such that $\O_{12}=-1$, we have the vielbein 
\begin{equation}
   V^{Aa} = \frac{1}{\sqrt{2}}
       \left( \begin{array}{cc} 
          u & \vb \\ v & -\ub 
       \end{array} \right)^{Aa}
\end{equation}
where we have introduced the one-forms
\begin{equation}
   u = \frac{d\xi}{\sqrt{V}} \qquad 
   v = \frac{1}{2V} \left( dV + id\s + \x d\xib - \xib d\x \right)
\end{equation}
and their complex conjugates $\ub$ and $\vb$. We find that the $SU(2)$
connection is given by
\begin{equation}
   {\o^A}_B = {\left( \begin{array}{cc} 
          \frac{1}{4}(v-\vb) & -u \\ \ub & -\frac{1}{4}(v-\vb)
        \end{array} \right)^A}_B
\end{equation}
while the $Sp(2)$ connection is 
\begin{equation}
   {\D^a}_b = {\left( \begin{array}{cc}
          -\frac{3}{4}(v-\vb) & 0 \\ 0 & \frac{3}{4}(v-\vb) 
        \end{array} \right)^a}_b \; .
\end{equation}
The triplet of K\"ahler forms is given by 
\begin{equation}
   {K^A}_B = {\left( \begin{array}{cc}
           \frac{1}{2}(u \wedge \ub - v \wedge \vb)  &  u \wedge \vb \\
           v \wedge \ub  &  - \frac{1}{2}(u \wedge \ub - v \wedge \vb)
         \end{array} \right)^A}_B \; .
\end{equation}
With these definitions, one finds that the coset space
$SU(2,1)/U(2)$, satisfies the conditions for a
quaternionic manifold. 

So far our discussion has ignored the most important aspect of the
hypermultiplet sigma-model. We note that the kinetic terms
in~\eqref{Hke} were in terms of a modified derivative~\eqref{covDer},
which included the gauge fields $\cA^i_\a$. It appears that the
hypermultiplet is charged under a $U(1)$ symmetry. Comparing with our
discussion of gauged supergravity given in Appendix B, we see that this
is indeed the case. The coset space $\cM_H$ admits an Abelian
isometry generated by the Killing vector
\begin{equation}
   k = \pt_\s = iV^{-1} \left( \pt_v - \pt_{\vb} \right) \; .
\end{equation}
In general we can write the modified derivative~\eqref{covDer} in the
covariant form~\eqref{covD}
\begin{equation}
   D_\a q^u = \pt_\a q^u + g \cA^i_\a k_i^u
\end{equation}
with
\begin{equation}
   g k_i = -2\e (x^{11})\a_i k = -2i\e (x^{11})\a_i V^{-1} \left( \pt_v -
           \pt_{\vb} \right)
\; .
\end{equation}
(Note that the gauge coupling is absorbed in $\a_i$.) For consistency,
the $k^i_u$ should be writable in terms of a triplet of prepotentials
given by~\eqref{cPcond}. This is indeed the case and we find the
prepotentials 
\begin{equation}
   g{{\cP_i}^A}_B = {\left( \begin{array}{cc} 
          - \frac{1}{4}i\e (x^{11})\a_i V^{-1} & 0 \\ 0 &
          \frac{1}{4}i\e (x^{11})\a_i V^{-1}
        \end{array} \right)^A}_B \; .
\end{equation}
Thus it appears that the $\s$-component of the hypermultiplet is
charged under each Abelian gauge field $\cA_\a^i$, with a charge
proportional to $\a_i$. In particular, we can write the covariant
derivative as
\begin{equation}
   D_\a \s = 
     \pt_\a \s +\frac{1}{4\sqrt{2}\p}\left(\frac{4\p}{\k_5}\right)^{2/3}
     \a_{\rm GUT}\,\e (x^{11})\b_i\cA^i_\a
\end{equation}
where $\b_i$ are integers characterizing the first Pontrjagin class of
the Calabi-Yau. 

If this interpretation is correct, the rest of the action should
coincide with the general form for gauged supergravity given in
Appendix B. It is clear that the vector multiplets are already in the
correct form. Comparing the bosonic action~\eqref{actparts} with the
general form~\eqref{kinetic}, we see that the gravitational and vector
kinetic terms exactly match. (In the Appendix we have set the
five-dimensional gravitational coupling $v/\k^2$ to unity.) The
structure of the metric $G_{ij}$ is identical, as is the appearance of
Chern-Simons couplings. The compactification gives an interpretation
of the numbers $d_{ijk}$ in the K\"ahler potential~\eqref{cKdef}
and~\eqref{kpot}. They are the Calabi-Yau intersection numbers. 

The final check of this identification is to calculate the form of
the potential. We have in general, from~\eqref{potential},
\begin{equation}
\begin{split}
   g^2V &= - 2 g^2 G_{ij} \tr \cP^i\cP^j + 4 g^2 b_i b_j \tr \cP^i\cP^j
          + \frac{g^2}{2} b^i b^j h_{uv} k^u_i k^v_j \\
     &= \frac{1}{4}V^{-2}G^{ij}\a_i\a_j \; ,
\end{split}
\end{equation}
exactly matching the derived potential. 

Thus we can conclude that the bulk effective action is described by a
set of Abelian vector multiplets coupled to a single charged
hypermultiplet. The vector sigma-model manifold $\cM_V$ has the
general form described in Appendix B, but now the $d_{ijk}$ in the
K\"ahler potential~\eqref{cKdef} have the interpretation as Calabi-Yau
intersection numbers. The hypermultiplet manifold $\cM_H$ is the coset
space $SU(2,1)/U(2)$. A $U(1)$ isometry, corresponding to
the shift symmetry of the dualized three-form, is gauged. The charge
of the hypermultiplet scalar field under each Abelian vector field
$A_\a^i$ is given by $\a_i$. 

\vspace{0.4cm}

The appearance of gauged supergravity when non-zero modes are included
has been seen before in the context of type II compactifications on
Calabi-Yau manifolds to four-dimensions~\cite{sp,michelson}. It is
natural to ask why this gauging arises. The appearance of a
potential term is easy to interpret. We have included a non-zero
four-form field strength $G_{IJKL}$ on four-cycles of the
Calabi-Yau. These contribute an energy proportional to the square of
the field strength. For fixed total charge $\a_i$ (the integral of $G$
over a cycle), the energy is reduced the larger the four-cycle. Thus
it is no longer true that all points in Calabi-Yau moduli space have
the same energy. As an example we see that the potential naturally
drives the Calabi-Yau to large volume, minimizing the $G^2$ energy. 

     From the five-dimensional point of view, once we have a potential
term, the theory must be gauged if it is to remain supersymmetric. We see
that it is the dual of the five-dimensional three-form which is
gauged. This arises because of the Chern-Simons term in eleven
dimensions. Turning on non-zero modes, this term acts as an electric
source for the five-dimensional three-form, though dependent on the
gauge fields $\cA^i$. Dualizing, the invariance $\s\to
\s+\mathrm{const}$ is a reflection of an absence of local electric
charge. Thus it not surprising that the effect of the electric
Chern-Simons terms is to modify this to a local gauge symmetry. We note
that from this argument it can only ever be the five-dimensional
three-form which becomes gauged by non-zero modes, whatever
particular compactification to a $N=1$ five-dimensional theory is
considered.

\vspace{0.4cm}

We end this section by giving the the specific form of the fermionic
supersymmetry variations. These are calculated using the general forms
given in~\eqref{psisusy},~\eqref{lsusy} and~\eqref{zsusy}, together
with the explicit expressions for the vielbein, connections, Killing
vectors and prepotentials given above. We find
\bsea
\d \psi_\a^A &=& \nabla_\a\e^A 
     + \frac{\sqrt{2}i}{8}
          \left({\g_\a}^{\b\g}-4\d_\a^\b\g^\g\right)b_i\cF_{\b\g}^i\e^A
     - {{P_\a}^A}_B \e^B \nn \\ && \qquad
     - \frac{\sqrt{2}}{12}V^{-1}b^i\a_i\g_\a\,\e (x^{11}){{\t_3}^A}_B\e^B 
     \\
\d \l^{xA} &=& b_i^x\left(-\frac{1}{2}i\g^\a\partial_\a b^i\e^A
             -\frac{1}{2\sqrt{2}}\g^{\a\b}\cF_{\a\b}^i\e^A
             -\frac{i}{2\sqrt{2}}V^{-1}\a^i\e (x^{11}){{\t_3}^A}_B\e^B
             \right) \\
\d \z^a &=& 
     - i {{Q_\a}^A}_B \g^\a \e^B
     - \frac{i}{\sqrt{2}}b^i\a_i V^{-1}\e (x^{11}){{\t_3}^a}_B\e^B
\label{susy5}
\esea
where $\t_i$ with  $i=1,2,3$ are the Pauli spin matrices and we have
the matrices
\begin{equation}
\begin{aligned}
   {{P_\a}^A}_B &= {\left( \begin{array}{cc} 
           \frac{\sqrt{2}i}{96}V \e_{\a\b\g\d\e}G^{\b\g\d\e} & 
           V^{-1/2}\pt_\a\x \\
           - V^{-1/2}\pt_\a{\bar\x} & 
           - \frac{\sqrt{2}i}{96}V \e_{\a\b\g\d\e}G^{\b\g\d\e}
        \end{array} \right)^A}_B \\
   {{Q_\a}^A}_B &= {\left( \begin{array}{cc}
           \frac{\sqrt{2}i}{48}V \e_{\a\b\g\d\e}G^{\b\g\d\e} 
               - \frac{1}{2}V^{-1}\pt_\a V &
           V^{-1/2}\pt_\a\x \\
           V^{-1/2}\pt_\a{\bar\x} & 
           \frac{\sqrt{2}i}{48}V \e_{\a\b\g\d\e}G^{\b\g\d\e} 
               + \frac{1}{2}V^{-1}\pt_\a V 
        \end{array} \right)^A}_B
\end{aligned}
\end{equation}

\section{The domain wall solution}

In this section, we would like to find the simplest BPS solutions
of the five--dimensional theory, including the coupling to
the potential terms induced by the nonzero mode. As we will
see, these solutions provide the appropriate background for a reduction to
four dimensions and can therefore be viewed as the ``vacua'' of the theory.
After a general derivation of the solutions we will discuss
several limiting cases of interest. In the next section one of these
limiting cases will be used to derive the four--dimensional effective
action to first nontrivial order.

\subsection{The general solution}

Let us first simplify the discussion somewhat by concentrating on the
fields which are essential. Since we would like to find solutions that
couple to the bulk potential terms we should certainly keep the
hypermultiplet scalar $V$ (the Calabi--Yau breathing mode) and the
vector multiplet scalars $b^i$ (the shape moduli). It turns out that those
fields plus the five--dimensional metric are already sufficient. The
action~\eqref{S5} can be consistently truncated to this reduced field
content leading to
\bea
 2\k_5^2 S_5 &=& - \int_{M_5}\sqrt{-g}\left[R+G_{ij}
         \partial_\a b^i\partial^\a b^j+\frac{1}{2}V^{-2}\partial_\a V
         \partial^\a V+\frac{1}{2}V^{-2}G^{ij}\a_i\a_j+\l (\cK - 6)
         \right] \nn\\
      && \qquad\qquad
         + 2\sqrt{2}\int_{M_4^{(1)}}\sqrt{-g}\, V^{-1}\a_ib^i - 2\sqrt{2}
         \int_{M_4^{(2)}}\sqrt{-g}\,V^{-1}\a_ib^i\; .
 \label{S5_red}
\eea
Note that we have explicitly added the Lagrange multiplier term which
ensures the constraint~\eqref{b_cons} on $b^i$.
For a finite Calabi--Yau volume $V$, that is, for an uncompactified
internal space, the potential terms in this action do not vanish and, hence,
flat space is not a solution of the theory. Therefore, the question arises
of what the ``vacuum'' state of the theory is. A clue is provided by the
fact that cosmological--type potentials in $D$ dimensions generally couple
to $D-2$ branes. This is well known from the eight--brane~\cite{8brane}
which appears as a solution of the massive extension of type IIA
supergravity~\cite{romans} in ten dimensions. There, the eight--brane
couples to a cosmological--type potential which consists of a single
``cosmological'' constant multiplied by a certain power of the dilaton.
A way to understand to appearance of an eight--brane in this context
is to dualize the cosmological constant to a nine--form antisymmetric
tensor field which, according to the usual counting, should couple
to an $8+1$--dimensional extended object. A systematic study of $D-2$
brane solutions in various dimensions using a generalized Scherk--Schwarz
reduction can be found in ref.~\cite{dom}. The present case is
somewhat more complicated in that it involves $h^{1,1}$ scalar fields (as
opposed to just the dilaton) and, correspondingly, $h^{1,1}$ constants
$\a_i$ (as opposed to just one cosmological constant). Still, we can take
a lead from the massive IIA example and dualize each of the constants
$\a_i$ to a four--form antisymmetric tensor field. This would leave us
with a theory that contains $h^{1,1}$ such antisymmetric tensor fields
and, hence, a corresponding number of different types of three--branes
that couple to those. The constants
$\a_i$ can then be identified as the charges of these different types
of three--branes. Since those constants are fixed in terms of the
underlying theory (and are generically nonzero) one
cannot really look for a ``pure'' solution which carries only one type
of charge. Instead, what we are looking for is a multi--charged three--brane
which is a mixture of the various different types as specified by
the charges $\a_i$. Clearly, the transverse space for a three--brane
in five--dimensions is just one--dimensional. Given that the boundary
source terms necessarily introduce dependence on the $x^{11}$ coordinate
this one--dimensional space can only be in the direction of the orbifold.

\vspace{0.4cm}

     From the above remarks it is now clear that the proper Ansatz for
the type of solutions we are looking for is given by
\bea
 ds_5^2 &=& a(y)^2dx^\m dx^\n\eta_{\m\n}+b(y)^2dy^2 \nn \\
 V &=& V(y) \label{3_ans}\\
 b^i &=& b^i(y)\nn\; ,
\eea
where we use $y=x^{11}$ from now on. The equations of motion
derived from the action~\eqref{S5_red} still contains the Lagrange
multiplier $\l$. It can be eliminated using
eqs.~\eqref{b_prop}--\eqref{bd_prop} from appendix A.
A solution to the resulting equations of the form~\eqref{3_ans} is still
somewhat hard to find, essentially due to the complication caused by the
inclusion of all $(1,1)$ moduli and the associated K\"ahler structure. The
trick is to express the solution in terms of certain functions $f^i=f^i(y)$
which are only implicitly defined rather than trying to find fully explicit
formulae. It turns out that those functions are fixed by the equations
\begin{equation}
 d_{ijk}f^jf^k=H_i\; ,\quad H_i=2\sqrt{2}k\a_i|y|+k_i\label{beta_def}
\end{equation}
where $k$ and $k_i$ are arbitrary constants. Then the solution can be written
as
\bea
 V &=&\left(\frac{1}{6}d_{ijk}f^if^jf^k\right)^2 \nn \\
 a &=&\tilde{k}V^{1/6}\nn\\
 b &=& kV^{2/3} \\
 b^i &=&V^{-1/6}f^i\nn
 \label{solution}
\eea
where $\tilde{k}$ is another arbitrary constant. We should check that
this solution is indeed a BPS state of the theory; that is, that it
preserves four of the eight supercharges. For the reduced field content,
the supersymmetry transformations~\eqref{susy5} lead to the following
Killing spinor equations
\bsea
 \d\psi_\m^A=0 & \quad :\quad &
    \g_\m\left(\frac{a'}{a}\g_{11}\e^A-\frac{\sqrt{2}b}{6V}
                 b^i\a_i\, \e (y)\, {{\t_3}^A}_B \e^B\right) = 0 \\
 \d\psi_{11}^A=0 & \quad :\quad &
    {\e^A}'-\frac{\sqrt{2}b}{12V}b^i\a_i\g_{11}\, \e (y) \,
                   {{\t_3}^A}_B \e^B = 0 \\
 \d\l^{xA}=0 & \quad :\quad &
    {b^i}'\g_{11}\e^A+\frac{b}{\sqrt{2}V}\left(
               \a^i-\frac{2}{3}b^j\a_jb^i\right)
               \e (y)\,{{\t_3}^A}_B \e^B = 0 \\
 \d\z^a = 0 & \quad :\quad &
    \frac{V'}{V}\g_{11}\e^A-\frac{\sqrt{2}b}{V}b^i\a_i\,\e (y)\,
              {{\t_3}^A}_B \e^B = 0\; ,
 \label{killing}
\esea
where the prime denotes the derivative with respect to $y$. These
equations are satisfied for the solution~\eqref{solution} if the
spinor $\e^A$ takes the form
\begin{equation}
 \e^A = a^{1/2}\e^A_0\; ,\quad \g_{11}\e^A_0 = {(\t_3)^A}_B \e^B_0\; ,
\end{equation}
where $\e^A_0$ is a constant spinor. As a result, the solution preserves
indeed four supercharges.

As can be seen from eq.~\eqref{beta_def} the solution is described in terms
of $h^{1,1}$ linear functions $H_i$. This follows the general pattern of
$p$--brane solutions coupled to $n$ different charges which can be
expressed in terms of $n$ harmonic functions on the transverse space. In our
case the number of charges $\a_i$ is precisely
$h^{1,1}$ and the transverse space is just one--dimensional leading to
linear functions. Generally, elementary brane solutions have singularities
at the location of the branes which have to be supported by brane worldvolume
theories. The pure bulk theory does not impose any restrictions on the
number and locations of these singularities. Correspondingly, if we would
just consider the bulk part of the action~\eqref{S5_red} we could place an
arbitrary number of  parallel three--branes anywhere on the orbifold.
However, the theory~\eqref{S5_red} involves two four--dimensional boundary
actions which provide source terms that should be matched. This is possible,
in the present case, because the height of the boundary potentials
in~\eqref{S5_red} is set by the three--brane charges $\a_i$. If we decide
that the solution should have no further singularities other than those
matched by the two boundaries we arrive at the specific form of the
harmonic functions $H_i$ in eq.~\eqref{beta_def}. In fact, we have
\begin{equation}
 {H_i}'' = 4\sqrt{2}k\a_i(\d (y)-\d (y-\p\r ))\; ,
\end{equation}
indicating sources at the orbifold planes $y=0,\p\r$. 
Recall that we have restricted the range of $y$ to $y\in [-\p\r ,\p\r ]$
with the endpoints identified. This explains the second delta--function
at $y=\p\r$ in the above equation.

In conclusion, the solution~\eqref{solution} represents a multi--charged
double domain wall (three--brane) solution with the two walls located
at the orbifold planes. It preserves four--dimensional Poincar\'e invariance
as well as four of the eight supercharges and has therefore the correct
properties to make contact with four--dimensional $N=1$ supergravity.
More precisely, those theories should arise as a dimensional reduction of
the five--dimensional theory on the domain wall background. In this sense,
the solution~\eqref{solution} can be viewed as the vacuum state of the
five--dimensional theory. In the next section we will carry out this
reduction explicitly to the first nontrivial order in the charges $\a_i$.
As a result, we will obtain the K\"ahler potential, the superpotential
and the gauge kinetic functions that specify the four--dimensional $N=1$
supergravity. From the perspective of the four--dimensional theory the domain
wall solution plays an interesting r\^ole. It is oriented precisely
in the four uncompactified dimensions and carries the physical gauge
and gauge matter fields. Therefore, at low energy four--dimensional space--time
gets identified with the three--brane worldvolume. In this sense, our
Universe lives on the worldvolume of a three--brane. Before we make this
point more explicit we would like to discuss some examples and limiting
cases of the general solution which will be useful in the following.

\subsection{Universal solution}

In ref.~\cite{losw} we have presented a related three--brane solution which
was less general in that it involved the universal Calabi--Yau
modulus $V$ only. Clearly, we should be able to recover this solution
from eq.~\eqref{solution} if we consider the specific case $h^{1,1}=1$.
Then we have $d_{111}=6$ and it follows from eq.~\eqref{beta_def} that
\begin{equation}
 f^1 = \left(\frac{\sqrt{2}}{3}k\a_1|y|+k_1\right)^{1/2}\; .
\end{equation}
Inserting this into eq.~\eqref{solution} provides us with the explicit
solution in this case which is given by
\bea
 a &=& a_0H^{1/2} \nn \\
 b &=& b_0H^2\qquad\qquad H=\frac{\sqrt{2}}{3}\a |y|+c_0\; ,\quad \a =\a^1 
 \label{u_sol}\\
 V &=& b_0H^3\nn \; .
\eea
The constant $a_0$, $b_0$ and $c_0$ are related to the integration constants
in eq.~\eqref{solution} by
\begin{equation}
 a_0=\tilde{k}k^{1/2}\; ,\qquad b_0=k^3\; ,\qquad c_0=\frac{k_1}{k}\; .
\end{equation}
Eq.~\eqref{u_sol} is indeed exactly the solution that was found in
ref.~\cite{losw}. It still represents a double domain wall. However, in
contrast to the general solution it couples to one charge $\a =\a^1$ only.
Geometrically, it describes a variation of the five--dimensional metric and
the Calabi--Yau volume across the orbifold. The form of the
solution~\eqref{u_sol} is typical for brane solutions that couple to one charge
and, in fact, fits into the general scheme of domain walls in various
dimensions~\cite{dom}.

\vspace{0.5cm}

One may ask if a structure as simple as the above universal solution is,
in some way, also part of the general solution~\eqref{solution} even
if $h^{1,1}>1$. To see that this is indeed the case, we define constants
$\bar{\a}^i$ and $\a$ by
\begin{equation}
 d_{ijk}\bar{\a}^j\bar{\a}^k=\frac{2}{3}\a_i\; ,\qquad 
 \a = 9\left(\frac{1}{6}d_{ijk}\bar{\a}^i\bar{\a}^j\bar{\a}^k\right)^{2/3}\; .
 \label{ab_def}
\end{equation}
In addition, we choose the following special values for the integration
constants $k_i$ in eq.~\eqref{beta_def}
\begin{equation}
 k_i=6kc_0\frac{\a_i}{\a}
\end{equation}
where $c_0$ is an arbitrary constant. Thanks to this specific choice, we
can easily solve~\eqref{beta_def} for $f^i$. Inserting the result into
eq.~\eqref{solution} gives the explicit solution
\bea
 a &=& a_0H^{1/2} \nn \\
 b &=& b_0H^2\; ,\qquad H=\frac{\sqrt{2}}{3}\a |y|+c_0 \label{u_sol1}\\
 V &=& b_0H^3\nn \\
 b^i &=& 3\a^{-1/2}\bar{\a}^i\nn\; .
\eea
As before, $a_0$ and $b_0$ are constants expressed in terms of the integration
constants in ~\eqref{solution} as
\begin{equation}
 a_0=\tilde{k}k^{1/2}\; ,\qquad b_0=k^3\; .
\end{equation}
Hence, for arbitrary values of $h^{1,1}$, we have identified a special case
of the general solution~\eqref{solution} where the fields $a$, $b$ and $V$
behave in exactly the same way as in the universal solution~\eqref{u_sol}.
The charge $\a$ which appears in this special solution is now a complicated
function of the various charges $\a_i$ in the way defined by
eq.~\eqref{ab_def}. In addition, the shape moduli $b^i$ are constant.
Consequently, for this special solution the metric and the Calabi--Yau
volume vary as in the universal solution while the shape of the Calabi--Yau
space is fixed.

\subsection{Another simple example}

A nontrivial example where the domain wall solution can be obtained
explicitly is provided by
\begin{equation}
 h^{1,1}=3\; ,\qquad d_{123}=1\; ,
\end{equation}
and $d_{ijk}=0$ otherwise. The K\"ahler potential is then given by
\begin{equation}
 \cK = 6\,b^1b^2b^3\label{STU} \; .
\end{equation}
In a four--dimensional effective theory the real fields $b^i$ are
promoted to complex scalars. Then the K\"ahler potential~\eqref{STU}
is associated with the coset space $\left[ SU(1,1)/U(1)\right]^3$
\cite{cfg} and describes the STU--model.
Due to the simple structure of intersection numbers eq.~\eqref{beta_def}
can be easily solved for the functions $f_i$ resulting in
\begin{equation}
 f^i  = (H_1H_2H_3)^{1/2}H_i^{-1}\; .
\end{equation}
Inserting into eq.~\eqref{solution} then gives the explicit solution
\bea
 V &=& (H_1H_2H_3)^{-1} \nn \\
 a &=& \tilde{k}(H_1H_2H_3)^{-1/6}\qquad\qquad H_i=2\sqrt{2}k\a_i |y|+k_i\\
 b &=& k(H_1H_2H_3)^{-2/3}\nn \\
 b^i  &=& (H_1H_2H_3)^{2/3}H_i^{-1}\nn
\eea
for $i=1,2,3$. As before $k$, $\tilde{k}$ and $k_i$ denote constants.

\subsection{The linearized solution}

In the next section we would like to calculate the four--dimensional
effective theory by reducing on the general domain wall
solution~\eqref{solution}. Clearly, to be able to do this we need an
explicit form for this solution. So far, we have obtained explicit
formulae only for specific examples. Unfortunately, the quadratic
equations~\eqref{beta_def} can generally not be solved exactly. However,
we can still try to find a sensible approximate solution. Since the
appearance of the domain walls is triggered by the charges $\a_i$ one
obvious approach is to find an approximate solution as an expansion in
$\a_i|y|$. This is what we will do in the present subsection. We will
restrict ourselves to the first nontrivial order; that is, to the terms
linear in $\a_i |y|$, although in principle one could continue the
procedure and find higher order terms as well. The result will be the
basis for the calculation of the four--dimensional effective action up
to order $\a_i$ in the next section.

\vspace{0.4cm}

Under which conditions does an expansion in $\a_i|y|$ actually make sense?
Inspection of the right hand side of eq.~\eqref{beta_def} shows that one
should require
\begin{equation}
 \left|\frac{k_i}{k}\right|\gg 2\sqrt{2}\p\r|\a_i| \label{cond1}
\end{equation}
for all $i=1,\cdots h^{1,1}$. In the following, we assume that the integration
constants $k_i$, $k$ are chosen such that these conditions are satisfied.
Eventually, those integration constants become functions of
the four--dimensional moduli. Then eq.~\eqref{cond1} represents a
constraints on the four--dimensional moduli space. We will return to this
point later.

In order to solve eq.~\eqref{beta_def} up to linear order in $\a_i |y|$
we start with the Ansatz
\begin{equation}
 f^i=A^i|y|+B^i \label{ans}
\end{equation}
with yet unknown constants $A^i$, $B^i$. Inserting this into~\eqref{beta_def}
leads to the relations
\begin{equation}
 k_i=d_{ijk}B^jB^k\; ,\qquad k\a_i = \frac{1}{\sqrt{2}}d_{ijk}A^jB^k
 \label{AB}
\end{equation}
which fix $A^i$ and $B^i$ in terms of the integration constants and the
charges. Before we explicitly calculate the solution we would like to
introduce a new set of integration constants $V_0$, $\hat{R}_0$, $b^i_0$
which is better adapted to their r\^ole as four--dimensional moduli.
This new set is defined by the conditions
\bea
 V\rightarrow V_0&,&a\rightarrow 1\nn \\
 b\rightarrow \hat{R}_0&,&b^i\rightarrow b_0^i\nn
\eea
for $\a_i\rightarrow 0$. Inserting the Ansatz~\eqref{ans} into
eq.~\eqref{solution} and comparing the $\a_i$ independent parts with
the above conditions it is easy to find the equations
\bea
 V_0 = B^2 &,& b_0^i = b^{-1/3}B^i \nn \\
 \hat{R}_0 = kB^{4/3} &,& \tilde{k}B^{1/3} = 1 \nn
\eea
with
\begin{equation}
 B=\frac{1}{6}d_{ijk}B^iB^jB^k
\end{equation}
which relate the old and the new integration constants. The final missing
piece of information necessary is an expression
for the constants $A^i$ in eq.~\eqref{ans}. They are fixed by the second
eq.~\eqref{AB}. Fortunately, this equation can be solved explicitly using
the relations~\eqref{b_prop} from appendix A. We find
\begin{equation}
 A^i=-\frac{V_0^{1/6}}{\sqrt{2}}\frac{\hat{R}_0}{V_0}(\a^i-b_0^j\a_jb_0^i)\; .
\end{equation}
Using these results we finally get for the solution to order $\a_i |y|$
\bea
 V &=& V_0\left( 1+\frac{\sqrt{2}\hat{R}_0}{V_0}b_0^i\a_i(|y|-\p\r /2)
       \right)\nn \\
 a &=& 1+\frac{\sqrt{2}\hat{R}_0}{6V_0}b_0^i\a_i(|y|-\p\r /2)\nn \\
 b &=& \hat{R}_0\left( 1+\frac{2\sqrt{2}\hat{R}_0}{3V_0}b_0^i\a_i
       \right) (|y|-\p\r /2) \label{app_sol}\\
 b^i &=& b_0^i-\frac{\hat{R}_0}{\sqrt{2}V_0}\left(\a^i-\frac{2}{3}
         b_0^j\a_jb_0^i\right)(|y|-\p\r /2)\nn\; .
\eea
Note that we have performed the shift $|y|\rightarrow |y|-\p\r /2$ which we
can always do by a suitable redefinition of the constants. The rationale for
this shift is that we would like the orbifold integral over the
corrections in~\eqref{app_sol} to vanish. Then the moduli $V_0$, $\hat{R}_0$
and $b_0^i$ equal the average values of the corresponding fields over the
orbifold, that is
\bsea
 \orbav{V}=V&,&\orbav{a}=1 \\
 \orbav{b}=\hat{R}_0&,&\orbav{b^i}=b^i_0\label{aver}
\esea
where the average is defined as
\begin{equation}
 \orbav{f} = \frac{1}{2\p\r}\int_{-\p\r}^{\p\r}dy f(y)
\end{equation}
for any function $f=f(y)$. As we will see in the next section (and has
been demonstrated for the universal case in ref.~\cite{low1}) with this
convention the moduli part of the four--dimensional K\"ahler potential
takes its usual form; that is, it is not affected by corrections of
order $\a_i$.

\vspace{0.4cm}

In terms of the new moduli, the condition~\eqref{cond1} for the validity of
the linear approximation can be written in the form
\begin{equation}
 \bar{\e}\ll \left|\frac{1}{4}d_{ijk}b_0^jb_0^k(\b_i)^{-1}\right|\; ,\qquad
  \bar{\e}=\e\frac{\hat{R}_0}{V_0}\label{cond2}
\end{equation}
where
\begin{equation}
 \e = \frac{1}{16}\left(\frac{4\p}{\k_5}\right)^{2/3}\r\,\a_{\rm GUT}\; ,
 \qquad \b_i=-\frac{1}{8\pi^2}\int_{C_i}\tr R\w R\; .\label{eps}
\end{equation}
Here we have used the definition~\eqref{alpha_def} of the charges $\a_i$.
Note that $\bar{\e}$ is a dimensionless quantity computed from the 
11--dimensional Newton constant, the Calabi--Yau volume and the orbifold
radius. This is the quantity discussed by Banks and
Dine~\footnote{This can be seen by expressing $\hat{R}_0$ in terms of the
orbifold radius $R_0$ as measured with the 11--dimensional Einstein frame
metric which gives $\hat{R}_0=R_0V_0^{1/3}$ and using
equation~\eqref{fdconst}.} and eq.~\eqref{cond2} is the generalization
of the validity condition discussed in their paper~\cite{bd}. As they
explain, for the ``physical'' values of $\k$, $\r\hat{R}_0$ and
$v^{2/3}V_0$ (that is, the values that match the Newton constant and
the grand unification coupling and scale) $\bar{\e}$ is of order one
and, hence, not a very good expansion parameter. Of course, in our
case things depend on more details like the topological values
$d_{ijk}$, $\b_i$ of the Calabi--Yau space and the values of the
Calabi--Yau shape moduli $b^i_0$. Nevertheless, it seems likely that
the approximation is not particularly good at the physical
point. Therefore it would be very desirable to go beyond the linear
approximation in order to get a reliable low energy theory. For the
purpose of the present paper, we assume that we are in a part of the
moduli space  where eq.~\eqref{cond2} holds.

\vspace{0.4cm}

How does the linearized solution~\eqref{app_sol} relate to Witten's
11--dimensional solution~\cite{w} which was also constructed up to
linear order in the charge $\a_i=O(\k^{2/3})$? The explicit form of this
11--dimensional solution was given in ref.~\cite{low1} as an expansion in 
terms of Calabi--Yau harmonics. Comparison with the expressions given
in this paper shows that eq.~\eqref{app_sol} coincides with the zero mode
part of the 11--dimensional solution. For the universal case, a similar
relation was demonstrated in ref.~\cite{losw}. Our general
solution~\eqref{solution} therefore represents an exact generalization of the
original linear solution which holds to all orders in $\k^{2/3}$. Of course
``exactness'' merely indicates that it is an exact solution of the low
energy effective theory which itself has higher-dimension correction
terms which have not been included. Nevertheless it is a tantalizing
question to which extend it can be used to calculate corrections to
the four--dimensional theory beyond the linear order.


\section{Four--dimensional physics from five dimensions}


As an application of the results presented so far we would now like
to calculate the effective four--dimensional theory which results from
the reduction on the domain--wall background to lowest nontrivial order.
This shows how the properties of the domain wall affect
physically relevant quantities in four dimensions such as the K\"ahler
potential and the gauge kinetic functions. 

\vspace{0.4cm}

In order to establish the correspondence between four-- and five--dimensional
fields let us first look at lowest order; that is, at the terms
independent on the domain wall charges $\a_i$. In this case, the bulk
zero modes simply coincide with the $Z_2$ even fields. Therefore, we have
\bea
 ds^2 &=& \hat{R}_0^{-1}g_{\m\n}dx^\m dx^\n +\hat{R}_0^2(dx^{11})^2 \nn \\
 V &=& V_0 \\
 b^i &=& b^i_0 \nn
\eea
where $g_{\m\n}$ is the four--dimensional Einstein frame metric.
Hence, the surviving fields are $h^{1,1}+1$ four--dimensional scalars
$\hat{R}_0$, $V_0$ and $b_0^i$ (recall that the $b^i$ satisfy the
constraint~\eqref{b_cons})which describe the orbifold radius, the
Calabi--Yau volume and the Calabi--Yau shape, respectively. Furthermore,
the following zero modes arise from the even components of the antisymmetric
tensor fields
\bea
 \cA^i_{11} = \c^i&,&\cF_{\m 11}^i=\partial_\m\c^i \nn \\
 C_{\m\n 11} = \frac{1}{6}B_{\m\n}&,&G_{\m\n\r 11}=H_{\m\n\r}=3\partial_{[\m}
               B_{\n\r ]}\; .
\eea
As a result, we have additional $h^{1,1}$ scalars $\c^i$ and the
two--form $B_{\m\n}$ with fields strength $H_{\m\n\r}$. The latter can
be dualized to a scalar $\s_0$ in the usual manner as
\begin{equation}
 H_{\m\n\r} = V_0^{-2}{\e_{\m\n\r}}^\s\partial_\s\s_0\; .
\end{equation}
How do these scalar fields fit into four--dimensional chiral multiplets?
A straightforward reduction to lowest order shows that the dilaton $S$
and the $h^{1,1}$ $T$--moduli $T^i$ should be defined as
\begin{equation}
 S=V_0+i\sqrt{2}\s_0\; ,\qquad T^i=\hat{R}_0b^i_0+i\sqrt{2}\c^i\; .\label{ST}
\end{equation}
In addition to those fields arising from the bulk we of course keep the
boundary gauge and gauge matter fields.

Our next goal is to incorporate corrections of order $\a_i$ into the
picture. The domain--wall solution to this order has been explicitly worked
out in the previous subsection. From eq.~\eqref{app_sol} we have
\bea
 ds^2 &=& (1+\tilde{a})\hat{R}_0^{-1}g_{\m\n}dx^\m dx^\n +\hat{R}_0^2
          (1+\tilde{\g})(dx^{11})^2 \nn \\
 V &=& V_0(1+\tilde{h}) \\
 b^i &=& b_0^i+\tilde{b}^i \nn
\eea
where all quantities with a tilde are of order $\a_i$. They are given by
\bea
 \tilde{a} &=& \frac{2\sqrt{2}\hat{R}_0}{3V_0}b_0^i\a_i (|y|-\p\r /2)
               \nn \\
 \tilde{\g} &=& 2\tilde{a} \nn \\
 \tilde{h} &=& \frac{3}{2}\tilde{a} \\
 \tilde{b}^i &=& -\frac{\hat{R}_0}{\sqrt{2}V_0}\left(\a^i
                 -\frac{2}{3}b_0^j\a_jb_0^i\right) (|y|-\p\r /2)\nn\; .
\eea
These equations already contain much of the essential information
necessary to compute the corrections of order $\a_i$. For a complete
reduction, however, one should deal with a number of additional complications.
First of all, the corrections to the metric given above induce corrections
to the zero modes of the antisymmetric tensor fields. Further corrections
appear if one considers gauge and gauge matter fields on the boundary
as we do in the present context. Those fields lead to boundary source
terms in the Bianchi identities as well as in the Einstein equation
which have to be carefully integrated out. In practise, this means
that one has to solve the Bianchi identities~\eqref{Bianchi5} and
the linearized Einstein equations with gauge and gauge matter sources.
This leads to correction terms in the fields $G_{\a\b\g\d}$, $\cF^i_{\a\b}$
and $X_\a$ as well as in the metric that involve gauge and gauge matter
fields. For the universal case, all these corrections have been
determined in ref.~\cite{low1}. Finally, if one wishes to include
gaugino condensates one should solve the Bianchi identity~\eqref{Bianchi5}
for $X_\a$ in the presence of the condensate as explained in the end of
section 3. In an, equivalent, 11--dimensional framework this has been done
in ref.~\cite{low2}. Here, we will not repeat the procedures
described in these papers but rather just state the final result.

\vspace{0.4cm}

The four--dimensional $N=1$ supergravity theory with chiral fields
$Y^\imath$ is specified by a K\"ahler potential $K=K(Y,\bar{Y})$, a
holomorphic superpotential $W=W(Y)$ and a holomorphic gauge kinetic
function $f=f(Y)$. To fix the normalizations let us state the relevant
terms in the bosonic part of the component action
\bea
 S &=& -\frac{1}{2\k_P^2}\int_{M_4}\sqrt{-g}R \nn \\
    && -\int_{M_4}\sqrt{-g}\left[ K_{\imath\bar{\jmath}}\partial_\m Y^\imath
       \partial^\m\bar{Y}^{\bar{\jmath}}+e^{\k_P^2K}
       \left( K^{\imath\bar{\jmath}}
       D_\imath W\overline{D_\jmath W}-3\k_P^2|W|^2\right)
       +\mbox{D-terms}\right]\nn \\
    && -\frac{1}{4g_{\rm GUT}^2}\int_{M_4}\sqrt{-g}\left[
          {\rm Re}f(Y){\rm tr}F^2 +{\rm Im}f(Y){\rm tr}F\tilde{F}
         \right] \; . \label{SUGRA}
\eea
Here $K_{\imath\bar{\jmath}}=\frac{\partial^2 K}{\partial Y^\imath
\partial\bar{Y}^{\bar{\jmath}}}$ is the K\"ahler metric and
$D_\imath W=\partial_\imath W+\frac{\partial K}{\partial Y^\imath}W$ is
the K\"ahler covariant derivative acting on the superpotential. The dual
field strength $\tilde{F}_{\m\n}$ is defined as
$\tilde{F}_{\m\n}=\frac{1}{2}\e_{\m\n\s\r}F^{\s\r}$.
The four--dimensional Planck constant $\k_P$ and the four--dimensional
gauge coupling $g_{\rm GUT}$ are defined in terms of the constants of
the underlying five-dimensional theory as
\bsea
 \k_P^2 &=& 8\p G_N = \frac{\k_5^2}{2\pi\rho} \\
 g_{\rm GUT}^2 &=& 4\p\a_{\rm GUT} \label{kpg} \; .
\esea
To have the kinetic terms of the matter fields normalized as in the
above action we should perform the rescaling
\begin{equation}
  C\rightarrow \frac{g_{\rm GUT}}{\sqrt{2}}C\; .
\end{equation}
Then we find for the K\"ahler potential
\begin{equation}
 K=-\k_P^{-2}\ln (S+\bar{S})+\k_P^{-2}K_T+Z_{ij}C^{ip}\bar{C}^j_p
\end{equation}
where
\bea
 K_T &=& -\ln\left(\frac{1}{6}d_{ijk}(T^i+\bar{T}^i)(T^j+\bar{T}^j)
         (T^k+\bar{T}^k)\right) \label{KT}\\
 Z_{ij} &=&\exp (-K_T/3)\left( K_{Tij}-\e\b_k
           \frac{1}{S+\bar{S}}\tilde{\G}_{Tij}^k\right)\label{Z}
\eea
and
\bea
 \tilde{\G}_{Tij}^k &=& \G_{Tij}^k+\frac{1}{3}K_{Tij}(T^k+\bar{T}^k) \\
 K_{Tij} &=& \frac{\partial^2K_T}{\partial T^i\partial\bar{T}^j}\; ,\qquad
 \G_{Tij}^k=K_T^{kl}\frac{\partial K_{Tjl}}{\partial T^i}\; .
\eea
The matter field superpotential is given by the usual expression
\begin{equation}
 W=\frac{\sqrt{2}}{3}g_{\rm GUT}d_{ijk}f_{pqr}C^{ip}C^{jq}C^{kr}
\end{equation}
and the gauge kinetic functions take the form
\bsea
 f^{(1)} &=& S+\e\b_iT^i \\
 f^{(2)} &=& S-\e\b_iT^i\; .\label{f}
\esea
In addition, if the gauginos of the hidden $E_8$ group condense one finds the
nonperturbative superpotential~\cite{low2}
\begin{equation}
 W_{\rm gaugino}\sim\exp\left[ -\frac{6\p}{b_0\a_{\rm GUT}}(S-\e\b_iT^i)
                    \right]\; .
\end{equation}
We recall that the range of the $(1,1)$ indices is 
$i,j,k,\cdots=1,\cdots ,h^{1,1}$. Therefore we have $h^{1,1}$ moduli $T^i$
and the same number of matter fields $C^{ip}$, each in the fundamental
representation ${\bf 27}$ of $E_6$, labeled by the indices
$p,q,r,\cdots = 1,\cdots ,27$. From eqs.~\eqref{alpha_def} and \eqref{eps}
the quantities $\e$ and $\b_i$ are defined by
\begin{equation}
 \e = \frac{1}{16}\left(\frac{4\p}{\k_5}\right)^{2/3}\r\,\a_{\rm GUT}\; ,
 \qquad \b_i= -\frac{1}{8\pi^2}\int_{C_i}\tr R\w R\; ,\label{eps1}
\end{equation}
where $C_i$ are the four--cycles of the Calabi--Yau space. Note that
$\e$ is a dimensionless quantity, while $\b_i$ are topological integers
that can be computed for a given Calabi--Yau manifold. The interpretation
of the moduli $S$, $T^i$ in terms of the underlying geometry is encoded in
the eqs.~\eqref{ST} and \eqref{aver}. All moduli fields are dimensionless
and they measure the form of the internal manifold relative to the
dimensionful quantities $v$ and $\r$ that we have introduced.

In the previous subsection we have discussed a validity condition for
the linear approximation that has been used to derive the above
four--dimensional theory. For convenience, we would now like to translate
this condition into four--dimensional language. From eq.~\eqref{cond2}
and the definition of the moduli~\eqref{ST} we find
\begin{equation}
 \left|(S+\bar{S})\frac{\partial K_T}{\partial T^i}\right|
    \gg \left|2\e\b_i\right|\; .
\end{equation}
The above four--dimensional theory is only valid in the region of moduli
space where these inequalities are satisfied for all $i$. A violation of
the conditions implies that corrections of quadratic and higher order
in $\e\b_i$ to the action can no longer be ignored.

\vspace{0.4cm}

The four--dimensional theory we have derived above is a generalization
of the results for the universal case, obtained in ref.~\cite{low1}
and ref.~\cite{hp,noy}, to the full $(1,1)$ sector of the theory. As such
it represents a new result within the M--theory context. The terms
independent of $\e\b_i$ coincide with the effective theory computed from
the weakly coupled heterotic string using geometrical methods~\cite{f,ft}
or conformal field theory methods~\cite{jan}. In addition, we have
two corrections of order $\e\b_i$, one to the matter field metric
$Z_{ij}$, eq.~\eqref{Z}, and a threshold correction to the gauge kinetic
functions~\eqref{f}. To our knowledge, a derivation of these additional
terms for the full $(1,1)$ sector from weakly coupled heterotic string
theory does not exist in the literature. This is understandable as the
correction terms of order $\e\b_i$ are tiny in the weakly coupled region of
the moduli space whereas they can be sizable or even of order one in the
strongly coupled region. For a careful comparison between the strong and weak
coupling limits of the heterotic string in relation to their four--dimensional
effective actions we refer the reader to ref.~\cite{low1}. At this point,
all we would like to demonstrate is that we recover the universal result
of
\cite{low1} for the case $h^{1,1}=1$. In this case, $d_{111}=6$ and we
set $T=T^1$ and
$\b=\b^1$. Then we find from eqs.~\eqref{KT} and \eqref{Z}
\begin{equation}
 K_T = -3\ln (T+\bar{T})\; ,\qquad Z_{T\bar{T}}=\frac{3}{T+\bar{T}}
       +\frac{\e\b}{S+\bar{S}}\; ,
\end{equation}
and for the gauge kinetic functions
\begin{equation}
 f^{(1)} = S+\e\b T\; ,\qquad f^{(2)}=S-\e\b T\; .
\end{equation}
This is indeed the result obtained in ref.~\cite{low1}.
The terms of order $\e\b$ in this universal theory have
important consequences for soft supersymmetry breaking terms as has been
pointed out in ref.~\cite{hp,ckm,low2}. Phenomenological implications
of the generalized result will be discussed elsewhere.

\vspace{0.4cm}

{\bf Acknowledgments} 
A.~L.~would like to thank Jens Erler for helpful discussions.
A.~L.~is supported in part by a fellowship from Deutsche
Forschungsgemeinschaft DFG. A.~L.~and B.~A.~O.~are supported in part by 
DOE under contract No. DE-AC02-76-ER-03071. D.~W.~is supported in part by
DOE under contract No. DE-FG02-91ER40671.


\clearpage
\appendix{\Large \bf Appendix}
\renewcommand{\theequation}{\Alph{section}.\arabic{equation}}
\setcounter{equation}{0}


\section{The Calabi--Yau K\"ahler moduli space}

In this appendix we are going to review the structure of the $(1,1)$ moduli
space of a Calabi--Yau three--fold as needed for the purpose of the present
paper. In addition, we collect some related formulae which are used
in our calculations. In the presentation, we are following
ref.~\cite{str,co,bcf}. 

\vspace{0.4cm}

We consider a Calabi--Yau three--fold $X$ with coordinates $x^A$,
metric $g_{AB}$ and volume $V$ given by~\footnote{Unlike in the main text,
we omit the dimensionful coordinate volume $v$ in this appendix for
simplicity.}
\begin{equation}
 V = \int_X\sqrt{g}\; .
\end{equation}
Its K\"ahler form $\o_{AB}$ is defined in terms of the metric as
\begin{equation}
 \o_{a\bar{b}} = ig_{a\bar{b}}\; .
\end{equation}
The cohomology group $H^{1,1}$ with dimension $h^{1,1}$ has a basis of
harmonic $(1,1)$ forms which we call $\{ \o_{iAB}\}$. Since the K\"ahler
form is similarly a harmonic $(1,1)$ form it can be expanded in terms of this
basis as
\begin{equation}
 \o_{AB} = a^i\o_{iAB}\label{oexp}
\end{equation}
where $a^i$ are the K\"ahler moduli. For the metric $G_{ij}(a)$ on the
K\"ahler moduli space one finds
\begin{equation}
 G_{ij}(a) = \frac{1}{2V}\int_X\o_i\w (*\o_j)\label{cymetric}
\end{equation}
To express this metric explicitly as a function of the moduli we introduce
the K\"ahler potential
\begin{equation}
 \cK (a) = \int_X\o\w\o\w\o\; .
\label{Kdef}
\end{equation}
Inserting the expansion~\eqref{oexp} it is easy to see that
\begin{equation}
 \cK (a) = d_{ijk}a^ia^ja^k
\end{equation}
where $d_{ijk}$ are the intersection numbers defined by
\begin{equation}
 d_{ijk} = \int_X\o_i\w\o_j\w\o_k\; .
\end{equation}
In addition, it is useful to define the quantities
\bea
 \cK_i(a) &=& \int_X\o_i\w\o\w\o = d_{ijk}a^ja^k \\
 \cK_{ij}(a) &=& \int_X\o_i\w\o_j\w\o = d_{ijk}a^k\; .
\eea
Then the metric~\eqref{cymetric} can be expressed in terms of the moduli as
\begin{equation}
 G_{ij}(a) = -\frac{1}{2}\frac{\partial}{\partial a^i}
             \frac{\partial}{\partial a^j}\ln\cK (a)
           =-3\left(\frac{\cK_{ij}(a)}{\cK (a)}-\frac{3}{2}\frac{\cK_i(a)
             \cK_j(a)}{\cK (a)^2}\right).
\label{Gdef}
\end{equation}
We note two useful properties of the metric which follow directly from
the above explicit form, namely
\begin{equation}
 G_{ij}(a)a^j=\frac{3}{2}\frac{\cK_i(a)}{\cK (a)}\; ,\qquad
 G_{ij}(a)a^ia^j=\frac{3}{2}\; .\label{prop}
\end{equation}
Let us also introduce the connection on the moduli space
\begin{equation}
 \G_{ij}^k(a) = \frac{1}{2}G^{kl}(a)\frac{\partial G_{ij}(a)}{\partial a^l}
\end{equation}
which reads explicitly
\begin{equation}
 \G_{ij}^k(a)=-\frac{3}{2}G^{kl}\left( \frac{d_{ijl}}{\cK (a)}
              -9\frac{\cK_{(i}(a)\cK_{jl)}(a)}{\cK (a)^2}
              +9\frac{\cK_i(a)\cK_j(a)\cK_l(a)}{\cK (a)^3}\right)\; .
\end{equation}
By straightforward computation one finds the simple property
\begin{equation}
 \G_{ij}^k(a)a^i = -\d^k_j\; .
\end{equation}
In the reduction from eleven to five dimensions we encounter a number
of integrals over the Calabi--Yau space which we would like to express
as functions of the moduli. We find for those integrals
\bea
 \int_X\sqrt{g} &=& V = \frac{1}{6}\cK (a) \nn \\
 \int_X\sqrt{g}\o_{ia\bar{b}}{\o_j}^{a\bar{b}} &=& 2VG_{ij}(a) \\
 \int_X\sqrt{g}g^{a\bar{b}}\o_{ia\bar{b}}g^{c\bar{d}}\o_{ic\bar{d}} &=&
     -2VG_{ij}(a)-\cK_{ij}(a)\nn \; .
\eea
In the five--dimensional effective theory that results from the reduction
on the Calabi--Yau space the volume modulus $V$ becomes part of the
universal hypermultiplet whereas the shape moduli fall into
vector--multiplets. It is therefore appropriate to scale out the volume $V$
and to define the shape moduli $b^i$ as
\begin{equation}
 b^i=V^{-1/3}a^i\; .
\end{equation}
Of course they are not independent but have to vary such that the volume
is unchanged. This is expressed by the constraint
\begin{equation}
 \cK (b)\equiv d_{ijk}b^ib^jb^k=6 \label{cons3}
\end{equation}
which follows directly from their definition and the fact that $V=\cK (a)/6$.
The metric for these shape moduli is then simply given by
\begin{equation}
 G_{ij}(b)=V^{2/3}G_{ij}(a)
\end{equation}
In the main text we will often drop the argument of the metric and
other related quantities. Whenever we do so it is to be understood
that we are referring to $G_{ij}(b)$. 
The other quantities introduced above scale in a simple way if
$a^i$ is replaced by $b^i$. Using those scalings and and the
constraint~\eqref{cons3} one can easily rewrite all equations in terms of
$b^i$. In particular, if we introduce the shape moduli $b_i$ with lowered
index defined by $b_i=G_{ij}(b)b^j$ the properties~\eqref{prop} can be
rewritten as
\begin{equation}
 b_i = \frac{1}{4}\cK_i(b)\; ,\qquad b_ib^i=\frac{3}{2}\; .\label{b_prop}
\end{equation}
If the moduli are viewed as fields in the five--dimensional uncompactified
space with coordinates $x^\a$ those equations lead the differential
equations
\bea
 b_i\partial_\a b^i &=& 0 \\
 b_i\left(\nabla_\a\nabla^\a b^i+\G_{jk}^i(b)\partial_\a b^i\partial^\a
     b^k\right) &=& 0\; ,\label{bd_prop}
\eea
which are useful to simplify the five--dimensional equations of motion.


\section{Einstein-Maxwell $D=5$, $N=1$ supergravity with gauged
hypermultiplets} 

The purpose of this appendix is to give a summary of the general form
of the coupling of matter and Abelian gauge fields to gravity in five
dimensions which preserves $N=1$ supersymmetry. There are three types
of multiplet: a gravitational multiplet with a graviton, two
gravitinos and a vector field; a vector multiplet with a vector field,
two gauginos and a real scalar; and a matter hypermultiplet with two
spinors and four scalar fields. The general coupling of Abelian
multiplets to gravity, but with none of the fields charged under the
Abelian symmetry, was first given by G\"unaydin \textit{et
al.}~\cite{GST1}. The same authors generalized this to the gauged case
where some of the fields became charged in a subsequent
paper~\cite{GST2}. The coupling to uncharged matter hypermultiplets
was introduced by Sierra~\cite{Sierra}. In the following we will
generalize to the case where the matter multiplets become charged,
giving the general gauging of the theory with Abelian vector
multiplets. We will not consider general gauging of a non-Abelian
vector multiplet sector. 

\subsection{Gamma matrices and symplectic spinors}

Let us start by defining our conventions for spinors in five
dimensions. A general spinor has four complex components. However,
the spinor and its complex conjugate can be arranged into a pair which
satisfies a twisted reality condition. Following~\cite{GST1}, we
define the gamma matrix algebra 
\begin{equation}
   \left\{ \g^m,\g^n \right\} = -2\, \eta^{mn} 
     = -2\, \mathrm{diag}\left(-++++\right)^{mn}
\end{equation}
If $C$ is the charge conjugation matrix, so that
${\g^m}^\mathrm{T}=C\g^mC^{-1}$, one finds $C^2=-1$ so that the
spinors cannot be taken to be real. However, suppose we have a
symplectic metric $\O_{ab}$ in a $2n$-dimensional vector space such
that we raise and lower indices by 
\begin{equation}
   v^a = \O^{ab}v_b \qquad v_a = v^b \O_{ba} \qquad 
   \O_{ab} = - \O_{ba}
\label{symplectic}
\end{equation}
One can then arrange $n$ independent spinors and their conjugates into
a $2n$-dimensional vector of symplectic spinors $\l_a$ satisfying a
reality condition 
\begin{equation}
   {\bar \l}^a \equiv {\l_a}^{\dag}\g_0 
      = \left( \O^{ab}\l_b \right)^\mathrm{T} C
\end{equation}
Here one is using the fact that $\O$ squares to minus one to cancel
the minus one in $C^2=-1$ and so provide a reality condition. 

The simplest supersymmetry is $N=1$ with a single supercharge. Writing
the charge as a pair of symplectic spinors we see that the
superalgebra must have a $Sp(2)=SU(2)$ R-symmetry automorphism group,
corresponding to rotations in the two-dimensional symplectic space. We
will denote these symplectic indices, in the fundamental of $SU(2)$, by
$A,B,C,\ldots$. The indices are raised and lowered following the
conventions of \eqref{symplectic} where we write $\e_{AB}$ for the
metric $\O_{ab}$ with $\e_{12}=\e^{12}=1$. 
   
\subsection{Coupling of gravitational and Abelian vector multiplets}

In this section we will describe the general ungauged coupling of
$n_V$ Abelian vector multiplets and the gravitational multiplet
following ref.~\cite{GST1}. From the vector multiplets we have $n_V$
real scalars which we will label $\phi^x$ while there are no scalars
in the gravitational multiplet. The general kinetic energy
of these scalars is described by a sigma-model where the scalars can
be interpreted as coordinates on some Riemannian manifold $\cM_V$. As
usual, the form of the manifold is restricted by supersymmetry, and
encodes all the information to describe the general coupling of the
vector and gravitational multiplets. In particular, generally
no independent function, which may provide a potential is available.
Elucidating the structure of this manifold will be the
main object of this section. 

First, though, we summarize the component fields of the multiplets. Aside
from the scalar $\phi^x$, the vector multiplets contain $2n_V$
gauginos and $n_V$ vector fields. The spinors $\l^{Ax}$ are
symplectic-real with respect to the $SU(2)$ R-symmetry labeled by $A$
and transform as vectors in the tangent space of $\cM_V$. The vector
fields $\cA_\a^{x}$ are also vectors in the tangent space. The
gravitational multiplet is given by a f\"unfbein ${e_\a}^m$ a pair of
gravitinos $\psi_\a^A$, symplectic-real under the $SU(2)$ R-symmetry,
and a graviphoton $\cA_\a$. We will find that the vector fields and
graviphoton naturally group together, so will often be labeled by
$\cA_\a^i$ where $i=0,1,\ldots,n_V$. 

The structure of $\cM_V$ is as follows. (Here we will use slightly
different normalizations from that of~\cite{GST1} in order to match the
form which naturally arises in the dimensional reduction of
eleven-dimensional supergravity on a Calabi-Yau.) One starts with a
$n_V+1$-dimensional space $\cC$ with coordinates $a^i$. A metric
$G_{ij}(a)$ on this space by  
\begin{equation}
   G_{ij}(a) = -\frac{1}{2}\partial_i\partial_j \cK(a)
\label{tG}
\end{equation}
where the partial derivative is with respect to the coordinates $a^i$
and $\cK$ is a homogeneous polynomial of degree three 
\begin{equation}
   \cK = d_{ijk}a^ia^ja^k
\end{equation}
(One notes that this metric becomes degenerate for certain values of
$a^i$.) One then imagines restricting to the space $K=6$. This
$n_V$-dimensional manifold with a metric induced from the pull-back of
the metric $G_{ij}(a)$ is the sigma-model manifold $\cM_V$. It is
precisely the same structure described by equations~\eqref{Kdef},
\eqref{Gdef} and \eqref{cons3} in the previous appendix, where the
metric was derived from compactification on a Calabi-Yau. However here
we have no geometrical interpretation of the symmetric tensor
$d_{ijk}$. 

Let us write $b^i$ for the values of the coordinates in $\cC$
restricted to the subspace $\cM_V$. Since the scalar fields $\phi^x$
are coordinates on $\cM_V$, we have $b^i=b^i(\phi^x)$. We can then
write the induced metric on $\cM_V$ as
\begin{equation}
   g_{xy} = b^i_x b^i_x G_{ij}
\label{gxydef}
\end{equation}
where $b^i_x=\partial b^i/\partial\phi^x$ and $G_{ij}$ is the metric
on $\cC$ evaluated at 
\begin{equation}
   \cK = d_{ijk} b^i b^j b^k = 6
\label{cKdef}
\end{equation}
(In general, when we drop the argument of $G_{ij}$ it is to be
understood as the restriction of the metric to $\cM_V$.)
We can also write the inverse matrix $G^{ij}$ in terms of the inverse
$g^{xy}$. One finds 
\begin{equation}
   G^{ij} = b^i_x b^i_x g^{xy} + \frac{2}{3}b^ib^j
\end{equation}
In the following we will adopt the convention that all $i,j,k$ indices
are raised and lowered with the metric $G_{ij}$ and $x,y,z$ indices
are raised and lowered with $g_{xy}$. There is a Levi-Civita
connection on $\cM_V$, giving the covariant derivative
\begin{equation}
   D_x v^y = \partial_x v^y + {\G^y}_{xz} v^z
\end{equation}
The corresponding Riemann curvature is highly constrained due to the
particular form of the metric $G_{ij}(a)$ given in \eqref{tG}. 

The full action and supersymmetry variations will be given at the end
of this appendix. Here let us simply summarize the transformation
properties of the component fields with respect to $\cM_V$. As we have
already mentioned, the $n_V$ scalar fields $\phi^x$ are interpreted as
coordinates on $\cM_V$. We will sometimes write them implicitly in
terms of $b^i$. The $n_V+1$ vector fields $\cA_\a^i$, including the
graviphoton, live naturally in the tangent space to $\cC$ at the
subspace $\cM_V$. Thus their indices are raised and lowered with
$G_{ij}$. Clearly the vectors with components in the $\cM_V$ tangent
space are the fields of the vector multiplet, while the graviphoton
$\cA_\a$ is given by the component orthogonal to $\cM_V$. Thus, since
$b_ib^i_x=0$, we have the decomposition %
\begin{equation}
   \cA_\a^x = b^x_i \cA^i_\a\; , \qquad \cA_\a = \frac{2}{3}b_i \cA^i_\a\; .
\end{equation}
The $2n_V$ symplectic gauginos $\l^{Ax}$ live in the tangent
space of $\cM_V$. In particular, the spacetime covariant derivative of
$\l^{Ax}$ must include a contribution from the Levi-Civita connection
on $\cM_V$,
\begin{equation}
   D_\a \l^{Ax} = \nabla_\a \l^{Ax} + \pt_\a\phi^y{\G^x}_{yz}\l^{Az}
\end{equation}
where $\nabla_\a$ is the conventional spacetime covariant derivative
with spin-connection. The graviton and gravitino are both scalars on
the manifold $\cM_V$. 

\subsection{Coupling of hypermultiplets}

We now turn to the form of the general coupling of uncharged
matter hypermultiplets to the Abelian-vector-gravity system described
above. This was first given by Sierra in~\cite{Sierra}. With $n_H$
hypermultiplets we have $4n_H$ scalar fields $q^u$ ($u=1,\ldots,4n_V$)
and $2n_H$ symplectic-real fermions $\z^a$ ($a=1,\ldots,2n_V$). Again the
coupling is fixed by giving the form of the scalar field
sigma-model. Fortunately one finds that the sigma model for the vector
multiplet and hypermultiplet scalars factorizes as $\cM_V\times\cM_H$,
with no cross couplings between the two types of field. The vector
multiplet sigma-model manifold remains exactly as described in the
previous section. Meanwhile, one finds, as for $N=2$ theories in four
dimensions, that the hypermultiplet scalars parameterize a quaternionic
manifold. As for the coupling of vector multiplets, there is in
general no independent function describing a potential, unless the
theory admits gauging. 

The structure of quaternionic geometry in supersymmetry was first
given in~\cite{BW}. Here we will follow the discussion
in~\cite{andetal}. The quaternionic manifold is a Riemannian manifold
with holonomy $Sp(2)\times Sp(2n_H)$. In the context of supersymmetry,
the $Sp(2)$ group is the $SU(2)$ R-symmetry. All tangent space
indices can thus be decomposed into the holonomy groups. Thus if
$q^u$ are coordinates on the manifold, we have, for instance, vielbein
one-forms 
\begin{equation}
   V^{Aa} = V^{Aa}_u dq^u
\end{equation}
where $A$ is an $SU(2)$ index and $a$ an index in the fundamental of
$Sp(2n_H)$. Let us write $\O_{ab}$ for the symplectic metric
preserved by $Sp(2n_H)$. The metric on the quaternionic manifold can
then be written as
\begin{equation}
   h_{uv} = V^{Aa}_uV^{Bb}_v\O_{ab}\e_{AB}
\end{equation}
In fact, more generally we have
\begin{equation}
\begin{aligned}
   V^{Aa}_u V^B_{va} + V^{Aa}_v V^B_{ua} &= h_{uv} \e^{AB} \\
   V^{Aa}_u V^b_{vA} + V^{Aa}_v V^b_{uA} &= 
      \frac{1}{n_H} h_{uv} \O^{ab}
\end{aligned}
\end{equation}
Since the holonomy has a product structure, the spin-connection
compatible with the metric also decomposes into an $SU(2)$ connection
and a $Sp(2n_H)$ connection. Let us write the corresponding one-forms
as ${\o^A}_B$ and ${\D^a}_b$. 

The metric on the quaternionic manifold is hermitian with respect to
any of three complex structures ${J^A}_B$ which fill out the adjoint
representation of $SU(2)$. As such they satisfy the quaternionic
algebra under matrix multiplication. There is correspondingly a
triplet of K\"ahler forms in the adjoint representation
\begin{equation}
   {K^A}_B = ({K^A}_B)_{uv}dq^u\wedge dq^v
\end{equation}
The K\"ahler forms are closed with respect to the $SU(2)$
connection. Thus, dropping the $SU(2)$ indices 
\begin{equation}
   dK + \o \wedge K = 0
\end{equation}
Finally one finds that the $SU(2)$ curvature is proportional to the
K\"ahler forms. Using the normalization of~\cite{andetal}, we have
\begin{equation}
   R = d\o + \o\wedge\o = - K
\end{equation}
One notes that in the rigid supersymmetry limit this condition becomes
$R=0$ and the quaternionic manifold becomes a hyper-K\"ahler
manifold. 

As in the case of the vector multiplet, we would like to describe the
transformation properties of the various component fields on
$\cM_H$. As mentioned above, the hypermultiplet scalar fields $q^u$
are coordinates on $\cM_H$. Unlike the gauginos in the vector
multiplet, one finds that the fermions $\z^a$ in the matter
multiplet are symplectic-real not with respect to the $SU(2)$ symmetry
but rather with respect to the $Sp(2n_H)$ symmetry, satisfying the
reality condition given in \eqref{symplectic}. From this we see that
$\z^a$ live in the $Sp(2n_H)$ part of the tangent
bundle. Consequently, as for the gauginos, there is a corresponding
correction piece in the spacetime covariant derivative from the
$Sp(2n_H)$ connection
\begin{equation}
   D_\a \z^a = \nabla_\a \z^a + \pt_\a q^u {{\D_u}^a}_b \z^b
\label{Deta}
\end{equation}
The bosonic fields in the vector and gravitational multiplets are all
scalars on $\cM_H$. However, one recalls that both the gauginos
$\l^{Ax}$ and the gravitinos $\psi_\a^A$ were doublets under
$SU(2)$. Thus they are in the $SU(2)$ part of the $\cM_H$ tangent
space. Consequently, their covariant derivatives get a correction,
when coupled to matter hypermultiplets, from the $SU(2)$ connection on
$\cM_H$. Namely we have
\begin{equation}
\begin{aligned}
   D_\a \l^{Ax} &= \nabla_\a \l^{Ax} + \pt_\a\phi^y{\G^x}_{yz}\l^{Az}
       + \pt_\a q^u {{\o_u}^A}_B \l^{Ax} \\
   D_\a \psi^A_\b &= \nabla_\a \psi^A_\b 
       + \pt_\a q^u {{\o_u}^A}_B \psi^A_\b
\end{aligned}
\label{Dpsilm}
\end{equation}

\subsection{Gauging the hypermultiplet sector}

We now turn to the question of how this structure is modified when the
matter fields become charged under the Abelian vector
symmetries. Again we will follow closely the description of the
four-dimensional case given in~\cite{andetal}. For the matter scalars
to be charged we must have an Abelian isometry of $\cM_H$ which can
become gauged. 

Suppose $\cM_H$ admits $n_V+1$ Abelian isometries (one for each
vector field). There is a Killing vector $k^u_i$ for each
isometry. If the isometries are to preserve the quaternionic structure
of $\cM_H$, each Killing vector can be written in terms of a
$SU(2)$ triplet of prepotentials ${{\cP_i}^A}_B$. These are sections
of the adjoint $SU(2)$-bundle on $\cM_H$ with derivatives related to
$k^u_i$. In general we have  
\begin{equation}
   k^u_iK_{uv} = \pt_v \cP_i + [\o_v,\cP_i]
\label{cPcond}
\end{equation}
so that in particular
\begin{equation}
   k^u_i = -\frac{1}{3}\tr K^{uv}\pt_v\cP_i
\end{equation}
where $\tr AB={A^A}_B{B^B}_A$. These prepotentials are the analogs of
the $D$-term prepotentials of $N=1$ supersymmetry. For Abelian
isometries, there is an ambiguity in the definition of $\cP_i$. They
may be shifted by constants matrices $\cC_i$ satisfying
$[\cC_i,\cC_j]=0$. These constants are the extension of
Fayet-Iliopoulos terms to $N=1$ $D=5$ supergravity. 

The procedure of gauging is now to make these isometry transformations
local functions in spacetime. The simplest modification is that the
spacetime derivative of the scalar fields $q^u$, coordinates in
$\cM_H$, becomes a gauge covariant derivative. Writing $g$ for the
general gauge coupling (this could be different for each Abelian
isometry) we have 
\begin{equation}
   \pt_\a q^u \ \longrightarrow\ 
      D_\a q^u = \pt_\a q^u + g \cA^i_\a k_i^u
\label{covD}
\end{equation}
However, because isometries in $\cM_H$ also produce a mapping between
tangent space vectors, we are in addition gauging both the $SU(2)$ and
the $Sp(2n_H)$ holonomy groups. Thus the corresponding connections in
the covariant derivatives are modified. Equations~\eqref{Deta} and
\eqref{Dpsilm} are modified to  
\begin{equation}
\begin{aligned}
   D_\a \z^a  &= \nabla_\a \z^a + \left( D_\a q^u {{\D_u}^a}_b 
       + g \cA_\a^i \pt_u k^v_i V^{uAa}V_{vAb} \right) \z^b \\
   D_\a \l^{Ax} &= \nabla_\a \l^{Ax} + \pt_\a\phi^y{\G^x}_{yz}\l^{Az}
       + \left( D_\a q^u {{\o_u}^A}_B 
       + g \cA_\a^i {{\cP_i}^A}_B \right) \l^{Bx} \\
   D_\a \psi^A_\b &= \nabla_\a \psi^A_\b 
       + \left( D_\a q^u {{\o_u}^A}_B 
       + g \cA_\a^i {{\cP_i}^A}_B \right) \psi^{Bx} 
\label{covDs}
\end{aligned}
\end{equation}

Perhaps the most important aspect of the gauging is that it introduces
a potential into the action. The potential is the analog of the
$D$-term potential in $N=1$, $D=4$ supergravity and is accompanied by
fermion mass terms. 

\subsection{The gauged action and supersymmetry transformations}

Having identified the structure of the sigma-models for the vector
multiplets and matter hypermultiplets, and discussed how the gauging of
charged matter is introduced, we finally turn to the form of the
gauged action and the supersymmetry transformations. 

The action can be split into kinetic terms, fermion mass terms,
potential and finally four-fermi terms which we will not give. We
write
\begin{equation}
   S = \int_{M_5} \sqrt{-g} \left( \cL_\mathrm{kinetic} 
         + \cL_\mathrm{four\ fermi}
         + \cL_\mathrm{fermi\ mass}
         - g^2 V \right)
\end{equation}
The kinetic terms are given by, with $\k_5=1$, 
\begin{equation}
\begin{aligned}
   \cL_\mathrm{kinetic} = -& \frac{1}{2}R
       - \frac{1}{2} G_{ij} \pt_\a b^i \pt^\a b^j
       - \frac{1}{2} G_{ij} \cF^i_{\a\b} \cF^{j\a\b}
       - \frac{1}{12\sqrt{2}} d_{ijk} \e^{\a\b\g\d\e} 
            \cA^i_\a \cF^j_{\b\g} \cF^k_{\d\e}
       - h_{uv} D_\a q^u D^\a q^v 
       \\ & 
       - \frac{1}{2} {\bar \psi}^A_\a \g^{\a\b\g} D_\b \psi_{\g A}
       - \frac{1}{2} {\bar \l^{Ax}} \g^\a D_\a \l_{Ax}
       - \frac{1}{2} {\bar \z^a} \g^\a D_\a \z_{a}
       \\ &
       + \frac{i}{4\sqrt{2}} \left( 
            {\bar\psi}^A_\g \g^{\a\b\g\d} \psi_{\d A}
            + 2 {\bar\psi}^{A\a} \psi_A^\b
            - {\bar\l^{Ax}} \g^{\a\b} \l_{Ax}
            - {\bar\z}^a \g^{\a\b} \z_{a} 
            \right) b_i \cF^i_{\a\b}
       \\ & 
       + \frac{1}{2\sqrt{2}} \left( 
            {\bar\l}^A_x \g^\a \g^{\b\g} \psi_{\a A}
            \right) b^x_i \cF^i_{\b\g}
       - \frac{i}{8\sqrt{2}} \left(
            {\bar\l^{Ax}} \g^{\a\b} \l^y_A
            \right) d_{ijk} b^i_x b^j_y \cF^i_{\a\b}
       \\ &
       - \frac{i}{2} \left(
            {\bar\l}^A_x \g^\a \g^\b \psi_{\a A}
            \right) b^x_i \pt_\b b^i
       + i \left( {\bar\z}_a \g^\a \g^\b \psi_{\a A} \right)
            V^{Aa}_u D_\b q^u
\end{aligned}
\label{kinetic}
\end{equation} 
Here $\cF^i=d\cA^i$ is the two-form Abelian field strength. All the
derivatives  $D_\a\l_{Ax}$ etc. are the full covariant derivatives
including gauging terms, as given in equations~\eqref{covD}
and~\eqref{covDs}. We see the appearance of the sigma-model metrics on
$\cM_V$ and $\cM_H$ in the scalar fields kinetic terms. We have
written the vector multiplet scalar fields in terms of the constrained
$b^i$ throughout. We could equally well have used the unconstrained
fields $\phi^x$. The sigma-model would then appear as
\begin{equation}
   \frac{1}{2} G_{ij} \pt_\a b^i \pt^\a b^j
      = \frac{1}{2} \left( G_{ij} b^i_x b^j_y \right) 
           \pt_\a\phi^x \pt^\a\phi^y
      = \frac{1}{2} g_{xy} \pt_\a\phi^x \pt^\a\phi^y
\end{equation}
where we have used equation~\eqref{gxydef}. One notes how the gauge
fields $\cA^i$ naturally couple to the $\cC$ metric $G_{ij}$, and in
addition that there is a Chern-Simons term $\cA\cF\cF$ depending on
the $d_{ijk}$ which define the $\cC$ metric. 

The fermion mass terms can be written as 
\begin{equation}
\begin{aligned}
   \cL_\mathrm{fermi\ mass} =& 
      - ig S^{AB} {\bar\psi}_{\a A} \g^{\a\b} \psi_{\b B}
      - g W^{AB}_x {\bar\l}^x_A \g^a \psi_{\a B}
      - g N_a^A {\bar\z}^a \g^\a \psi_{\a A}
      \\ &
      - ig M_{AxBy} {\bar\l}_{xA} \l_{yB}
      - ig M^{aAx} {\bar\z}_a \l_{Ax}
      - ig M^{ab} {\bar\z}_a \z_b
\end{aligned}
\label{fermimass}
\end{equation}
where all these terms depend on the Killing vectors and prepotentials
\begin{equation}
\begin{aligned}
   S_{AB} &= \frac{1}{\sqrt{2}} b_i {\cP^i}_{AB} \\
   W^x_{AB} &= - \sqrt{2} b^x_i {\cP^i}_{AB} \\
   N^{Aa} &= - \frac{1}{\sqrt{2}} V^{Aa}_u b^i k_i^u \\
   M_{AxBy} &= 
      - \frac{3}{\sqrt{2}} d_{ijk} b^{ix} b^{jx} {\cP^k}_{AB}
      - 3\sqrt{2} b^i_x b^j_y G_{ij} \left( b_k {\cP^k}_{AB} \right) \\
   M^{aAx} &= - \frac{1}{\sqrt{2}} V^{Aa}_u b^{ix} k^u_i \\
   M^{ab} &= \frac{1}{4\sqrt{2}} V^{Aa}_u V^{Bb}_v \e_{AB}
     \left( b^i \nabla^{[u} k^{v]}_i \right)
\end{aligned}
\end{equation}
Here $\nabla^{u} k^{v}_i$ is the covariant derivative of the Killing
vector on the $\cM_H$ quaternionic manifold. 

Finally we have the potential 
\begin{equation}
   V = - 2 G_{ij} \tr \cP^i\cP^j + 4 b_i b_j \tr \cP^i\cP^j
          + \frac{1}{2} b^i b^j h_{uv} k^u_i k^v_j
\label{potential}
\end{equation}
where $\tr AB={A^A}_B{B^B}_A$. Clearly the potential vanishes if there
is no gauging. We also note that it is possible to have a contribution
to the potential from pure Fayet-Iliopoulos terms. It is specified by
$k^u_i=0$ and $\cP^i=$const which satisfies eq.~\eqref{cPcond}. 

To complete the theory we also list the supersymmetry
transformations. The supersymmetry parameter $\e^A$ is an $SU(2)$
symplectic real spinor. The fermion variations
\begin{align}
   \d \psi_\a^A &= D_\a \e^A 
      - \frac{i}{6\sqrt{2}} \left(
          {\g_\a}^{\b\g} - 4 {\d_\a}^\b \g^\g 
          \right) b_i \cF^i_{\b\g} \e^A 
      + \frac{2ig}{3} S^{AB} \g_\a \e_B
      \label{psisusy} \\
   \d \l^{Ax} &= b^x_i \left(
          \frac{i}{2} \g^\a \pt_\a b^i 
          + \frac{1}{2\sqrt{2}} \g^{\a\b} \cF^i_{\a\b} 
          \right) \e^A + g W^{xAB} \e_B
      \label{lsusy} \\
   \d \z^a &= - i V^{Aa}_u \g^\a D_\a q^u \e_A - g N^{aA} \e_A
      \label{zsusy}
\end{align}
include corrections from the gauging. The bosonic variations on
the other hand receive no corrections, and read
\begin{align}
   \d {e_\a}^m &= \frac{1}{2} {\bar\e}^A \g^m \psi_{\a A}
      \label{esusy} \\
   \d \cA^i_\a &= \frac{1}{2\sqrt{2}} b^i_x {\bar\e}_A \g_\a \l^{Ax} 
      - \frac{i}{2\sqrt{2}} b^i {\bar\psi}^A_\a \e_A
      \label{cAsusy} \\
   \d b^i &= - \frac{i}{2} b^i_x {\bar\e}_A \l^{Ax}
      \label{bsusy} \\
   \d q^u &= \frac{i}{2} V^u_{Aa} {\bar\e}^A \z^a
      \label{qsusy}
\end{align}
This completes the description of the general coupling of Abelian
vector multiplets to charged matter in $D=5$ supergravity.



\end{document}